\documentclass[11pt]{article}

\usepackage{amsmath}
\usepackage{amssymb}
\usepackage{amsfonts}
\usepackage{latexsym}
\usepackage{graphicx}
\usepackage{latexsym}
\usepackage{color}

\usepackage{indentfirst}

\def\ln{\,\mbox{ln}\,}

\def\tr{\,\mbox{tr}\,}
\def\Tr{\,\mbox{Tr}\,}

\def\diag{\,\mbox{diag}\,}

\def\al{\alpha}
\def\be{\beta}
\def\ga{\gamma}
\def\Ga{\Gamma}
\def\de{\delta}
\def\De{\Delta}

\def\vp{\varepsilon}

\def\ka{\kappa}
\def\la{\lambda}

\def\si{\sigma}

\def\ph{\varphi}

\def\na{\nabla}
\def\pa{\partial}

\renewcommand{\vec}[1]{{\bf #1}}

\DeclareMathOperator{\cx}{\square}

\def\beq{\begin{eqnarray}}
\def\eeq{\end{eqnarray}}
\newcommand{\eq}[1]{(\ref{#1})}

\newcommand{\nn}{\nonumber}

\begin{document}

 \begin{center}

{\Large 
\bf Universal leading
 quantum correction to the Newton potential 
}

 \vskip 6mm

\textbf{Tib\'{e}rio de Paula Netto}$^a$
\footnote{E-mail: \ tiberio@sustech.edu.cn},
\textbf{Leonardo Modesto}$^a$
\footnote{E-mail: \ lmodesto@sustech.edu.cn},
and
\textbf{Ilya~L.~Shapiro}$^b$
\footnote{Also at Tomsk State Pedagogical University,
e-mail: \ ilyashapiro2003@ufjf.br}
%

\vskip 4mm

(a) Department of Physics,
\ Southern University of Science and Technology,
\\ Shenzhen, \ 518055, \ China
\vskip 2mm

(b) Departamento de F\'{\i}sica, \ ICE, \
Universidade Federal de Juiz de Fora
\\ Juiz de Fora, \ 36036-900, \ MG, \ Brazil
\vskip 2mm

\end{center}
\vskip 6mm

\begin{abstract}

\noindent
The derivation of effective quantum gravity corrections to Newton's
potential is an important step in the whole effective quantum field
theory approach. We hereby add new strong arguments in favor of
omitting all the diagrams with internal lines of the massive sources,
and we also recalculate the corrections to the Newtonian potential
using functional methods in an arbitrary parametrization of the
quantum fluctuations of the metric. The general proof of the gauge-
and parametrization-independence within this approach is also
explicitly given. On top of that, we argue that the universality of
the result  holds regardless of the details of the ultraviolet
completion of quantum gravity theory. Indeed, it turns out that the
logarithm quantum correction depends only on the low energy
spectrum of the theory that is responsible for the analytic properties
of loop's amplitudes.
\vskip 3mm


\end{abstract}

\vskip 2mm

\noindent


\section{Introduction}

Quantum corrections to Newton's potential for the classical
gravitational force is a kind of theoretical ``standard candle''
for the effective quantum gravity theory. The first calculation of these
corrections are half a century old \cite{Iwa}, but the explosion of
interest to the particular example and the effective quantum gravity
in general, started from the seminal works due to Donoghue
\cite{don94-PRL,don94}.
The total amount of publications (see, e.g.,
\cite{KirKhri,BjerrumBohr-2002} and \cite{don-reviews} and
references therein) is very large, such that we are probably not
aware of all of them. However, in our opinion, there remain at least
two relevant issues which were not sufficiently well discussed.

First, starting from the paper \cite{SQED} there was a growing
understanding that the gauge-fixing independence of the quantum
corrections is an important and useful criterion for the correctness
and consistent definition of these corrections. In the mentioned
paper, the simplified model of scalar electrodynamics was used
to prove that disregarding some of the diagrams is inconsistent as
this procedure does not provide the gauge-fixing independence of
the potential. In the subsequent papers, i.e.,
\cite{KirKhri,BjerrumBohr-2002}, all the diagrams were taken into
account and the invariance was proved (see, e.g., \cite{Kazakov1999}).
However, it turns out that there are serious reasons to take out
some of these diagrams \cite{DM97,Polemic}.

The calculations in most of the mentioned (and many other) papers
were done using Feynman diagrams where two massive scalar fields
model the massive bodies. Then, the complete set of diagrams
includes those of $L$- and $P$-type. The first, $L$-type subset
consists of the graphs with internal lines of massless gravitons and
massless gauge (Faddeev-Popov) ghosts. These diagrams give only
a very small ${\mathcal O}(r^{-3})$-type addition to the Newton
potential between the two-point masses. Phenomenologically, these
graphs may be irrelevant, but conceptually they are the most
important ones since the much stronger contributions coming
from the $P$-type diagrams should be disregarded because of
physical reasons \cite{Polemic}. The point is that our final purpose
is to get the quantum corrections to the force between two
\textit{macroscopic} bodies, which are modeled by the scalar field.
Taking into account the \textit{internal} lines of these scalars means
that we are regarding the massive macroscopic bodies as quantum objects.
For example, if the purpose is to explore the quantum effects on the motion
of an asteroid around the Sun, we are quantizing this asteroid and
the Sun. Or we are quantizing, e.g.,  the Earth and the Moon. Indeed,
the Moon does not like to be quantized. Thus, we may commit a
serious mistake in this way.

Two other arguments for omitting the $P$-type diagrams are as follows.
The macroscopic body is made from atoms, and most of the
mass of the atom nucleons is due to the quantum effects. Thus, the
argument concerning the expansions in $\hbar$ is not operational
in this case. Furthermore, nowadays there are strong experimental
confirmations of not quantizing the classical source in a physically
similar situation. It is good to remember the success of the
theoretical description of Casimir force (see, e.g.,
\cite{Casimir-BKMM}). Despite the existing controversies (see,
for example, the critical discussion in \cite{Casimir-Lamoreaux}),
there are no many doubts that the Casimir effect is experimentally
detected, and we know that its theoretical description does not
involve the quantization of the sources (such as conducting plates)
but only the electromagnetic field. There are no reasons to
assume that things should be different for the gravitational Casimir
effect because the basic formalism is, in general, similar
\cite{Quach2015}. Thus, the correct procedure is to separate and
use only $L$-type diagrams, i.e., those with only massless internal
lines. From the quantum field theory viewpoint, this means
that the object of our interest is the functional integral
\beq
e^{i\Ga}
\,=\,
\int dg \,d{\bar c} \,dc \,\exp\big\{iS_{grav}(g) + iS_{gf}(g)
+ iS_{gh}(g,{\bar c},c) + iS_{sources}(g,\Phi)\big\},
\label{pathint}
\eeq
where $g=g_{\al\be}$ is the quantum metric, ${\bar c}={\bar c}_\al$
and $c=c^\be$ are gauge ghosts, and $\Phi$ represent the external
massive sources.
The approach described above was pursued in important work
\cite{Duff74} and, most recently, in \cite{DM97}.
Following the last reference, in the calculations
described below, we choose these sources as a massive point-like
mass and a light point-like test particle, but in principle, the integral
(\ref{pathint}) can be explored for arbitrary massive sources, e.g.,
for the dark matter (DM), finite-size stars, or interstellar gas. In all
these cases, the integration over $\Phi$ is not included for the
reasons described above. This means that the variables $\Phi$
play the role of external parameters, independent of the
quantum variables such as the metric components.

A general observation is in order. There are relevant aspects of the
semiclassical approach (when metric is a classical background while
matter fields quantum), related to the possible transformations which
are mixing
classical and quantum variables \cite{Duff-preprint1980}. One can
try to elaborate the same idea in the \textit{anti-semiclassical}
approach, when only gravity is quantized. However, in the effective
framework, in the IR only the massless fields are relevant, leaving
only the effects of graviton and photon \cite{don94-PRL}. In this
effective setting, our \textit{anti-semiclassical} approach emerges
naturally and the mixing of the massless degree of freedom with the
ones of massive (macroscopic) degrees of freedom describing the
classical sources does not look reasonable. 

In practice, one can use the background field method for the useful
definition and derivation of the integral (\ref{pathint}). Then, the
effective action in the \textit{l.h.s.} is $\Ga(g,\Phi)$ and the
integral variables change accordingly. We shall discuss the explicit
formulas below.

The paper is organized in the following way. Sec.~\ref{sec2}
describes the derivation of the relevant part of $\Ga(g,\Phi)$.
Sec.~\ref{sec3} aims to define the gravitational field in the
Newtonian approximation, in the theory with quantum effective
contributions. Sec.~\ref{sec4} describes the motion of test particles
and defined the final form of the corrected Newtonian potential.
Sec.~\ref{sec5} is devoted to the general proof of the gauge-
and parametrization-independence of the effective quantum
correction to Newton's potential.
In Sec.~\ref{secLM} we elaborate on the constrains that an effective field theory should satisfy in order to be consistent in the quantum field theory framework, and after that we explain the universality of the leading correction to Newton's potential, which only depends on the fundamental theory low-energy spectrum.
Finally, in Sec.~\ref{sec6}
we draw our conclusions.

The notations include the Minkowski metric
$\eta_{\mu\nu} = \diag (1,-1,-1,-1)$, while the Riemann curvature
tensor is defined as
\beq
\label{Rie}
R^\al_{\,\cdot\, \be\mu\nu} \,=\,
\pa_\mu \Ga^\al_{\be \nu} - \pa_\nu \Ga^\al_{\be \mu}
+ \Ga^ \al_{\mu \tau} \, \Ga^\tau_{\be \nu}
- \Ga^\al_{\nu \tau } \, \Ga^\tau _{\be \mu}.
\eeq
The definitions of the Ricci tensor and the scalar curvature are
$R_{\mu\nu} = R^\al_. {}_{\mu\al\nu}$ and
$R = g^{\mu\nu} R_{\mu\nu}$, respectively. We also adopt
the units system with $c = 1$ and $\hbar  = 1$.

\section{Derivation of effective action}
\label{sec2}

In the effective quantum gravity \cite{don94,don-reviews}, it is
assumed that the low-energy effects are described by Einstein's
general relativity with zero cosmological constant because
this term is irrelevant at the astrophysical scale where the
Newtonian limit applies. The possible contributions of higher
derivative terms either decouple (see, e.g., \cite{324} for
preliminary analysis, coherent with the standard approach in
particle physics \cite{manohar,Ilisie}) or regarded as small
perturbations by definition~\cite{Simon-90}. The reader can
consult \cite{Burgess2003} and \cite{OUP} for a general
discussion of different approaches. However, the net result is
that we have to  start from the classical action of the form
\beq
S = - \frac{1}{\ka^2} \int d^4 x \sqrt{-g} \, R
+ S_{\rm sources}(g,\Phi),
\label{classical_action}
\eeq
where $\ka^2 = 16 \pi G$ and $G$ is the Newton constant. As the
source term, we shall consider the action of the free massive
particle $S_M$, with coordinates
$y^\mu$
\beq
S_M = - M \int ds
= - M \, \int \sqrt{g_{\mu\nu} \, d y^\mu d y^\nu}.
\label{actM}
\eeq

To quantize the gravitational field, we apply the background field
method with the most general (at the one-loop level) parametrization
of the metric, as formulated in Ref.~\cite{JDG-QG} (see
also~\cite{BTS-Vil-1}),
namely
\beq
\label{gen-bfm}
g_{\mu\nu}
&
\longrightarrow
&
g_{\mu\nu}^\prime \,=\, e^{2 \ga_0 \ka\si} \big[ g_{\mu\nu}
+ \ka ( \ga_1 \phi_{\mu\nu} + \ga_2 \phi  g_{\mu\nu})
\nn
\\
&&
\qquad \quad
+ \, \ka^2 (  \ga_3 \phi_{\mu\rho} \phi^\rho_\nu
+ \ga_4 g_{\mu\nu} \phi_{\rho\si} \phi^{\rho\si}
+ \gamma_5 \phi \phi_{\mu\nu}
+ \gamma_6 g_{\mu\nu} \phi^2) \big].
\eeq
In this formula, $g_{\mu\nu}$ is the background field and $\si$,
$\phi_{\mu\nu}$ are quantum fields. The indices are lowered and
raised with the background metric $g_{\mu\nu}$. Also,
$\phi = g_{\mu\nu} \phi^{\mu\nu}$. Finally, $\ga_{0,1,\ldots,6}$
are arbitrary coefficients which parameterize quantum
variables.

Treating the conformal factor $\si$ as a new field introduces an
artificial conformal symmetry \cite{FirPei}. In order to remove the
corresponding degeneracy, we implement the conformal gauge fixing
\beq
\label{conformal_gauge}
\si = \la \phi \, ,
\eeq
where $\la$ is a gauge-fixing parameter. Using \eq{conformal_gauge}
and expanding the exponential, one can rewrite Eq.~\eq{gen-bfm} in
the form
\begin{equation}
g'_{\mu\nu} = g_{\mu\nu}
+ \ka A_{\mu\nu}^{\al\be} \phi_{\al\be}
+ \ka^2 \phi_{\la\tau} B_{\mu\nu}^{\la\tau,\rho\si} \phi_{\rho\si}
+ {\mathcal O}(\ka^3),
\end{equation}
where
\begin{equation}
A_{\mu\nu}^{\al\be}
= \ga_1 \de_{\,\,\,\,\,\mu\nu}^{\al\be}
+ (\ga_2 + 2 \gamma_0 \la) g_{\mu\nu} g^{\al\be} \, ,
\end{equation}
and
\begin{equation}
\begin{split}
B_{\la\tau}^{\mu\nu,\al\be} = &\,\,
\frac{\ga_3}{4} ( \de_{\,\,\,\,\,\la\tau}^{\mu\al} g^{\nu\be}
+ \de_{\,\,\,\,\,\la\tau}^{\nu\al} g^{\mu\be}
+ \de_{\,\,\,\,\,\la\tau}^{\mu\be} g^{\nu\al}
+ \de_{\,\,\,\,\,\la\tau}^{\nu\be} g^{\mu\al} )
+ \ga_4 \de^{\mu\nu,\al\be} g_{\la\tau}
\\
&
+ \frac{1}{2} (\ga_5 + 2 \gamma_0 \la \ga_1)
( \de^{\mu\nu}_{\,\,\,\,\,\la\tau} g^{\al\be}
+  \de^{\al\be}_{\,\,\,\,\,\la\tau} g^{\mu\nu})
+ (\ga_6 + 2 \gamma_0 \la \ga_2 + 2 \gamma_0^2 \la^2)
g^{\mu\nu} g^{\al\be} g_{\la\tau} \, ,
\end{split}
\end{equation}
with
\begin{equation}
\de^{\mu\nu}_{\,\,\,\,\,\al\be}
= \frac12 (\de^\mu_\al \de^\nu_\be + \de^\mu_\be \de^\nu_\al)
\,
.
\end{equation}

To fix the diffeomorphism invariance and simplify the
calculations\footnote{The non-minimal gauge was considered
in \cite{DM97} and proved to give a gauge-independent result for the one-loop
contribution to Newton's gravitational potential, so we avoid
discussing it here.
},
we introduce the following minimal gauge fixing action \cite{JDG-QG},
\beq
S_{gf} = - \frac12 \int d^4 x \sqrt{-g} \,
\big[
\ga_1 \na_\la \phi_{\mu}^\la
- \tfrac12 \left(\ga_1 + 2 \ga_2 + 4 \gamma_0 \la \right)
\na_\mu \phi
\big]^2.
\label{mingauge}
\eeq
Performing the expansion of \eq{gen-bfm} in \eq{classical_action}
with the minimal gauge-fixing term, we arrive at the bilinear, in
the quantum metric, part of the action
\beq
\label{bili_phi}
(S + S_{gf})^{(2)} = \frac12 \int d^4 x \sqrt{-g} \,
\phi_{\mu\nu} \, {\bf H}^{\mu\nu,\al\be} \, \phi_{\al\be},
\eeq
where
\beq
&& \label{Hessian_1}
{\bf H}^{\mu\nu,\al\be} =
- ({\bf K}^{\mu\nu,\al\be} \Box
+ {\bf \Pi}^{\mu\nu,\al\be}
+ {\bf M}^{\mu\nu,\al\be}) \,,
\eeq
with
\beq
&&
{\bf K}^{\mu\nu,\al\be}
= A^{\mu\nu}_{\la\tau} \, K^{\la\tau,\rho\si}  A^{\al\be}_{\rho\si} ,
\nn
\\
&&
{\bf \Pi}^{\mu\nu,\al\be}
= A^{\mu\nu}_{\la\tau} \, K^{\la\tau,\rho\si} A^{\al\be}_{\rho\si}
- 2 B_{\la\tau}^{\mu\nu,\al\be} G^{\mu\nu},
\nn
\\
&&
{\bf M}^{\mu\nu,\al\be}
= \ka^2 (B_{\la\tau}^{\mu\nu,\al\be} T^{\la\tau}
- A^{\mu\nu}_{\la\tau} \, M^{\la\tau,\rho\si}  A^{\al\be}_{\rho\si}),
\label{K_parametrization}
\eeq
where we used the notations:
\beq
&&
G^{\mu\nu} = R^{\mu\nu} - \frac12 \, g^{\mu\nu} R,
\qquad
K^{\mu \nu, \al\be}
= \frac12 ( \de^{\mu \nu, \al\be} - \tfrac12 g^{\mu\nu} g^{\al\be} ),
\nn
\\
&&
\Pi^{\mu\nu,\al\be}  \,=\,
\frac12 (R^{\mu\al\nu\be} + R^{\mu\be\nu\al} )
+ \frac14 (g^{\mu\al} R^{\nu\be} + g^{\mu\be} R^{\nu\al}
+ g^{\nu\al} R^{\mu\be} + g^{\nu\be} R^{\mu\al})
\nn
\\
&&
\qquad
\qquad
\quad
-\, \frac12 (g^{\mu\nu} R^{\al\be} + g^{\al\be} R^{\mu\nu})
- K^{\mu\nu,\al\be} R.
\eeq
Furthermore, in the source sector,
\beq
\label{EMT}
&&
T^{\mu\nu} (x) = M \int ds \,
\de \left(x-y(s) \right) u^\mu u^\nu
\eeq
is the point-like mass energy-momentum tensor and
\beq
\label{M}
&&
M^{\mu\nu,\al\be} (x) =  \frac{M}{4} \int ds \,
\de \left(x-y(s) \right) u^\mu u^\nu u^\al u^\be.
\eeq
It is worth noting that the boldface notation are used for the
operators depending on the parametrization and gauge fixing.
Also, in \eq{EMT} and \eq{M},
\beq
u^\mu = \frac{dy^\mu}{ds}
\eeq
is the particle four-velocity and the delta
function is introduced through the relation
\beq
\int d^4 x \sqrt{-g(x)} \,
\de \left( x - y(s) \right) = 1.
\eeq

The standard Schwinger-DeWitt technique for the one-loop
divergences requires to reduce the Hessian \eq{Hessian_1}
to the standard expression for the minimal operator, i.e.,
$\hat{1} \Box + \hat{\bf P} - (1/6) \hat{1} R$.
Therefore, we multiply \eq{Hessian_1} with the negative
inverse of \eq{K_parametrization}, namely
\beq
 {\bf K}_{\mu\nu,\al\be}^{-1}
= (A^{-1})_{\mu\nu}^{\la\tau}
\, K_{\la\tau,\rho\si}^{-1}  (A^{-1})_{\al\be}^{\rho\si},
\eeq
where
\beq
K_{\la\tau,\rho\si}^{-1}
= 2 ( \de_{\la\tau,\rho\si} - \tfrac12 g_{\la\tau} g_{\rho\si} ),
\qquad
(A^{-1})_{\mu\nu}^{\al\be}
= \frac{1}{\ga_1} \Big( \de_{\,\,\,\,\,\mu\nu}^{\al\be}
- \frac{\ga_2 + 2 \gamma_0 \la}{\ga_1 + 4 \ga_2 + 8 \gamma_0 \la} \,  g_{\mu\nu}g^{\al\be} \Big).
\eeq
Adopting the suppressed index notation of Ref.~\cite{bavi85}, we get
\begin{equation}
\hat{\bf H} \equiv
- {\bf K}_{\mu\nu,\la\tau}^{-1} \cdot {\bf H}^{\la\tau,\al\be}
= \hat{1} \Box + \hat{\bf P} - \frac16 \, \hat{1} R,
\end{equation}
where $\hat{1} = \de_{\,\,\,\,\,\mu\nu}^{\al\be}$ is the identity matrix
for symmetric rank-2 tensors and
\begin{equation}
\hat{\bf P} =
{\bf K}_{\mu\nu,\la\tau}^{-1} \cdot {\bf \Pi}^{\la\tau,\al\be}
+ {\bf K}_{\mu\nu,\la\tau}^{-1} \cdot {\bf M}^{\la\tau,\al\be}
+ \frac16 \, \hat{1} R.
\end{equation}

Performing the substitution \eq{gen-bfm} in \eq{pathint},
we have the one-loop contribution to the effective action
\begin{equation}
\label{EA}
\bar{\Ga}^{(1)} = \frac{i}{2}  \Tr \ln \hat{\bf H}
- i  \Tr \ln \hat{\bf H}_{gh},
\end{equation}
where $\hat{\bf H}_{gh}$ is the bilinear part of the
Faddeev-Popov ghost action. The contribution of this term
to \eq{EA} is the same as in Ref.~\cite{JDG-QG} hence
we do not repeat it here.

The divergent part of the first term in \eq{EA} can be evaluated
by the trace of the coincidence limit of the Schwinger-DeWitt
expansion $\hat{a}_2$ coefficient
\cite{BDW-65} (we do not include here the surface terms, as
they are irrelevant for our purpose)
\begin{equation}
\label{a2}
\frac{i}{2} \Tr \ln \hat{\bf H}\,  \big|_{div} =
- \frac{\mu^{n-4}}{(4\pi)^2(n-4)} \int d^n x \sqrt{-g} \,
\tr \left\{\frac{1}{180} \hat{1} (R_{\mu\nu\al\be}^2 - R_{\mu\nu}^2)
+ \frac12 \hat{\bf P}^2
+ \frac{1}{12} \hat{\cal R}^2_{\mu\nu}
\right\},
\end{equation}
where
\begin{equation}
\hat{{\cal R}}_{\la\tau} = -
\frac12 (\de^\al_\mu R^\be_{\,\cdot\,\nu\rho\si}
+ \de^\be_\mu R^\al_{\,\cdot\,\nu\rho\si}
+ \de^\al_\nu R^\be_{\,\cdot\,\mu\rho\si}
+ \de^\be_\nu R^\al_{\,\cdot\,\mu\rho\si} )
\end{equation}
and $\mu$ is the renormalization group scale.

To simplify \eq{a2}, we use the identity
$g_{\mu\nu} u^\mu u^\nu = 1$ which implies in the following
equations for the expressions (\ref{EMT}) and (\ref{M}):
\beq
\label{traceM}
g_{\mu\nu} M^{\mu\nu,\al\be} = \frac14 \, T^{\al\be},
\eeq
\beq
M_{\mu\nu,\al\be} M^{\mu\nu,\al\be} = \frac{1}{16} \, T_{\mu\nu} T^{\mu\nu},
\eeq
\beq
\label{part_T_square}
T_{\mu\nu} T^{\mu\nu} = T^2,
\eeq
where
\beq
T \equiv T^\mu_\mu = M \int ds \,
\de \left(x-y(s) \right)
\eeq
is the trace of the energy-momentum tensor for a massive point-like
particle. There are also equations analogous to~\eq{traceM} for
other index contractions since $M^{\mu\nu,\al\be}$ is totally symmetric.

The evaluation of \eq{a2} follows the scheme explained in
detail in~\cite{JDG-QG}. Thus, skipping the algebra, we get, for the
one-loop divergences
\beq
\bar{\Ga}^{(1)}_{div}
&=&
- \frac{\mu^{n-4}}{(4\pi)^2(n-4)} \int d^n x \sqrt{-g} \,
\big\{
c_1 R_{\mu\nu\al\be}^2 + c_2 R_{\mu\nu}^2 + c_3 R^2
\nn
\\
&&
\quad
+ \,\,
\ka^2 c_4 R_{\mu\nu} T^{\mu\nu}
+ \ka^2 c_5 R T
+ \ka^4 c_6 T^2
\big\},
\label{divs-1}
\eeq
where the dimensionless coefficients are
given by
\begin{align}
& c_1 = \frac{53}{45},
\qquad
c_2 = -\frac{361}{90}
-\frac{4 \xi_2}{\ga_1^4 Z^2},
\qquad
c_3 = \frac{43}{36} + \frac{\xi_3}{3 \ga_1^4 Z^4},
\nn \\ &
c_4 =  - \frac{2 \xi_4}{ \ga_1^4 Z^4},
\qquad
c_5 = \frac{5}{24} - \frac{\xi_5}{6 \ga_1^4 Z^4},
\qquad
c_6 = \frac{1}{32} - \frac{\xi_6}{4 \ga_1^4 Z^4},
\label{c123456}
\end{align}
where $Z=\ga_1+ 4 \ga_2 + 8 \gamma_0 \la$, and the bulky expressions
for the parametrization dependent coefficients $\xi_i$ can be found in Appendix A.

In the limit $\ga_1 \to 1$, $\ga_{2,\ldots,6} \to 0$
and $\ga_0 \to 0$ for the parameters of (\ref{gen-bfm}),
Eq.~\eq{divs-1} reproduces the result of Ref.~\cite{DM97},
evaluated in the standard parametrization in the minimal gauge.

Using the classical equations of motion
\beq
R_{\mu\nu} - \frac12 g_{\mu\nu} R
= \frac{\ka^2}{2}  T_{\mu\nu}
\label{onshell}
\eeq
and \eq{part_T_square}, Eq.~\eq{EA} boils down to
\beq
\bar{\Ga}_{div}^{(1)} \big|_{ \text{on-shell} } =
- \frac{\mu^{n-4}}{(4\pi)^2(n-4)} \int d^n x \sqrt{-g}
\left\{
\frac{53}{45} R_{\mu\nu \alpha \beta}^2
- \frac{373 }{480}\, \ka^4 T^2
\right\},
\eeq
independently of the choice of parametrization and gauge
parameters, exactly as it should be.

To consider the Newtonian limit it is useful to
rewrite Eq.~\eq{divs-1} in a more convenient Weyl basis, using the
relations
\beq
R_{\mu\nu\al\be}^2 = 2 C^2 - E + \frac{1}{3}\,R^2,
\qquad
R_{\mu\nu}^2 = \frac12\, C^2 - \frac12\, E + \frac{1}{3}\,R^2.
\eeq
Here $C^2 = R_{\mu\nu\al\be}^2 - 2 R_{\mu\nu}^2 + 1/3 \, R^2\,$
is the square of the Weyl tensor and
$E = R_{\mu\nu\al\be}^2 - 4 R_{\mu\nu}^2 + R^2\,$ is the integrand
of the (topological at $n=4$) Gauss-Bonnet term. Discarding the
non essential for us topological term, we get for \eq{divs-1}:
\begin{equation}
\begin{split}
\label{div_Weyl}
\bar{\Ga}^{(1)}_{div}
= - \frac{\mu^{n-4}}{n-4} \int d^n x \sqrt{-g} \, \Big\{
\beta_2 C^2 - \frac13 \beta_0 R^2
-  2 \ka^2 \beta_{\scriptscriptstyle RT1} R_{\mu\nu} T^{\mu\nu}
+  \ka^2\beta_{\scriptscriptstyle RT2} R T
+ \ka^4 \beta_{\scriptscriptstyle TT} T^2
\Big\},
\end{split}
\end{equation}
with the following relations between the coefficients,
\beq
&&
\beta_2 = \frac{4 c_1 + c_2}{2(4 \pi )^2},
\qquad
\beta_0 = - \frac{c_1 + c_2 + 3 c_3}{(4 \pi )^2},
\label{beta1}
\\
&&
\beta_{\scriptscriptstyle RT1} = - \frac{c_4 }{2(4 \pi )^2},
\qquad
\beta_{\scriptscriptstyle RT2} =  \frac{ c_5 }{(4 \pi )^2},
\qquad
\beta_{\scriptscriptstyle TT} =  \frac{ c_6 }{(4 \pi )^2}.
\label{beta5}
\eeq

To reconstruct the effective action we use the rule (see, e.g.,
\cite{OUP} for the introduction and further references, including
the generalizations for the loops of massive fields)
\beq
\frac{\mu^{n-4}}{n-4}
\,\,\,
\longrightarrow
\,\,\,
\frac{1}{2} \ln \Big( \frac{\Box}{\mu^2} \Big),
\eeq
that is valid for the massless fields contributions such as the graviton.
In this way, we finally get
\begin{equation}
\begin{split}
\label{EA-Weyl}
\bar{\Ga}^{(1)} = &
- \int d^4 x \sqrt{-g} \,  \left\{
\frac12 \beta_2 C_{\mu\nu\al\be}
\ln \Big( \frac{\Box}{\mu^2} \Big) C^{\mu\nu\al\be}
- \frac{1}{6} \beta_0 R \ln \Big( \frac{\Box}{\mu^2} \Big) R
\right.
\\ &
\left.
-  \ka^2 \beta_{\scriptscriptstyle RT1}  R_{\mu\nu} \ln \Big( \frac{\Box}{\mu^2} \Big) T^{\mu\nu}
+ \frac12  \ka^2 \beta_{\scriptscriptstyle RT2} \, R \ln \Big( \frac{\Box}{\mu^2} \Big) T
+ \frac12  \ka^4 \beta_{\scriptscriptstyle TT} T \ln \Big( \frac{\Box}{\mu^2} \Big) T
\right\}.
\end{split}
\end{equation}

\section{Gravitational field in the Newtonian limit with quantum contributions}
\label{sec3}

Here we are going to obtain the gravitational field in the Newtonian
limit taking into account one-loop quantum corrections.
The effective action is given by the classical action $S$,
defined in \eq{classical_action}, and the one-loop correction
(\ref{EA-Weyl}), i.e.,
\beq
\label{finite_EA}
\Ga = S + \bar{\Ga}^{(1)}.
\eeq
To obtain the metric potentials with logarithm quantum corrections,
we follow Refs.~\cite{Burzilla:2020bkx,Burzilla:2020utr}
(see also~\cite{Modesto:2014eta,Giacchini:2018gxp,Giacchini:2018wlf}).
First, we consider small perturbations of the background metric around
the Minkowiski spacetime,
\begin{equation}
\label{background}
g_{\mu\nu} = \eta_{\mu\nu} + \ka h_{\mu\nu},
\qquad \qquad
\vert \ka h_{\mu\nu} \vert \ll 1
\end{equation}
and expand the action \eq{finite_EA} to
the relevant order in the weak-field approximation.
In this way, we get
\begin{equation}
\begin{split}
\label{HD-weak}
\Ga = & - \frac{1}{2}  \int d^4 x 
\left\{
\frac12 h_{\mu\nu} f_2 (\Box) \Box h^{\mu\nu}
- h^{\mu\nu} f_2 (\Box) \pa_\mu \pa_\la  h^{\la}_{\nu}
- \frac16 \, h \left[f_2 (\Box) + 2f_0 (\Box) \right] \Box h
\right.
\\
& \left.
+ \, \frac13 h
\left[f_2 (\Box) + 2f_0 (\Box) \right] \pa_\mu \pa_\nu h^{\mu\nu}
+ \frac13 h_{\al\be} [ f_2 (\Box) - f_0 (\Box) ]
\frac{\pa^\al \pa^\be \pa^\mu \pa^\nu}{\Box} h_{\mu\nu}
\right\}
\\
&
- \,\frac{\ka}{2} \int d^4 x \, \Big\{    h_{\mu\nu} f_{\scriptscriptstyle RT1} (\Box) T^{\mu\nu}
+  \ka^2 \beta_{\scriptscriptstyle RT2} \,
h_{\mu\nu} \ln \Big( \frac{\Box}{\mu^2} \Big) ( \pa^\mu \pa^\nu - \eta^{\mu\nu} \Box ) T
\Big\},
\end{split}
\end{equation}
where we defined the form factors
\begin{equation}
f_i (\Box) \,\,=\,\, 1 + \ka^2 \beta_{i}
\ln \Big(\frac{\Box}{\mu^2}\Big) \Box,
\qquad \qquad
i = 2,0, RT1.
\end{equation}
Let us note that in \eq{HD-weak},
$\Box = \eta^{\mu\nu} \pa_\mu \pa_\nu$ is the flat space
d'Alembertian. Also, the term proportional to $\beta_{\scriptscriptstyle TT}$ is
discarded here because it is $\propto M^2$ and, therefore, is beyond
the weak-field approximation.

Then, taking the functional derivative with
respect to $h_{\mu\nu}$ we get the equations
of motion
\begin{equation}
\begin{split}
\label{EOMHD}
& \hspace{-0.35cm}
f_2 (\Box) ( \Box h^{\mu\nu} - \pa^\mu \pa^\la h^\nu_\la
- \pa^\nu \pa^\la h^\mu_\la )
+ \frac13 [ f_2 (\Box) + 2 f_0 (\Box) ]
 [ \eta^{\mu\nu} (\pa^\al \pa^\be h_{\al\be} - \Box h) + \pa^\mu \pa^\nu h]
\\
& \hspace{-0.35cm}
+ \frac23 [ f_2 (\Box) - f_0 (\Box) ] \, \frac{\pa^\mu \pa^\nu \pa^\al \pa^\be}
{\Box} h_{\al\be}
= -\ka \Big[ f_{\scriptscriptstyle RT1} (\Box) T^{\mu\nu}
+  \ka^2 \beta_{\scriptscriptstyle RT2} \, \ln \Big( \frac{\Box}{\mu^2} \Big)
(\pa^\mu \pa^\nu  - \eta^{\mu\nu} \Box ) T
\Big].
\end{split}
\end{equation}

The second step is to consider the perturbation in the isotropic
Newtonian form
\beq
\label{h00-hij}
\ka h_{00} = 2 \ph (r),
\qquad
\ka h_{ij} = 2 \de_{ij} \psi (r)
\eeq
and choose, as a source of the gravitational potentials, a static
point-like mass located at the origin. In this case,
$u^\mu = \de^\mu_0$, and Eq.~\eq{EMT} becomes
\begin{equation}
\label{EMT_static}
T^{\mu\nu} = \de^\mu_0 \de^\nu_0 \,\rho,
\quad \quad
\rho = M \de (\vec{r}).
\end{equation}
The metric potentials~\eq{h00-hij} can be obtained from the 00--component
and the trace of Eq.~\eq{EOMHD}, which are given
respectively by
\beq
&&
\hspace{-1.5cm}
\big[ f_2(-\De) - f_0 (-\De) \big] \De \ph
 +
\big[ f_2(-\De) + 2 f_0 (-\De) \big]  \De \psi =
\nn
\\
&&
\hspace{3cm} =
\frac{3 \ka^2}{4} \Big[ f_{\scriptscriptstyle RT1} (-\De) + \ka^2
\beta_{\scriptscriptstyle RT2}  \ln \left( - \frac{\De}{\mu^2} \right) \! \De \Big] \rho,
\label{00}
\\
\label{trace}
&&
\hspace{-1.5cm}
f_0 (-\De) ( \De \ph - 2 \De \psi )
\,=\, - \frac{\ka^2}{4} \left[ f_{\scriptscriptstyle RT1} (-\De)
+ 3 \ka^2 \beta_{\scriptscriptstyle RT2} \ln ( - \De / \mu^2 ) \right] \rho,
\eeq
where we traded $\Box$ for $- \De$ since the metric is static.

To solve \eq{00} and \eq{trace} we perform the  loop expansion
of the potentials,
\begin{align}
&
\ph = \ph^{(0)}  + \ph^{(1)}  + {\mathcal O}(\hbar^2),
\\
&
\psi = \psi^{(0)}  + \psi^{(1)}  + {\mathcal O}(\hbar^2),
\end{align}
where $\ph^{(l)}$ and $\psi^{(l)}$ are of the order
${\mathcal O}(\hbar^l)$. Since  $\beta_i = {\mathcal O}(\hbar)$,
we get the equations at zero and first orders in $\hbar$ in the form
\beq
\label{0-order}
&&
\De \ph^{(0)} = \De \psi^{(0)} = \frac{\ka^2 M}{4} \de (\vec{r}),
\\
&&
\De \psi^{(1)}
 \,=\,
\frac{ \ka^2 \beta_2}{3}
\ln \Big( - \frac{\De}{\mu^2} \Big)  \De^2 (\ph^{(0)} + \psi^{(0)})
- \frac{\ka^2 \beta_0}{3} \ln \Big( - \frac{\De}{\mu^2} \Big)  \De^2 (\ph^{(0)} - 2 \psi^{(0)})
\nn
\\
&&
\qquad
\qquad
+ \,\,\frac{\ka^4 M}{4} ( -\beta_{\scriptscriptstyle RT1}  + \beta_{\scriptscriptstyle RT2})
\ln \Big( - \frac{\De}{\mu^2} \Big) \De \de (\vec{r}),
\label{00-loop}
\\
&&
\De\ph^{(1)} - 2 \De \psi^{(1)}
\,\,=\,\, \ka^2 \beta_0
\ln \Big( - \frac{\De}{\mu^2} \Big) \De^2 (\ph^{(0)} - 2 \psi^{(0)})
\nn
\\
&&
\qquad
\qquad
+ \,\, \frac{\ka^4 M}{4} ( \beta_{\scriptscriptstyle RT1} - 3 \beta_{\scriptscriptstyle RT2} )
\ln \Big( - \frac{\De}{\mu^2} \Big) \De \de (\vec{r}).
\label{trace-loop}
\eeq
In the three-dimensional Fourier space, we obtain the following
solutions for the transformed potentials:
\beq
&&
\ph^{(0)} (k) = \psi^{(0)} (k) = - \frac{\ka^2 M}{4 k^2},
\nn
\\
&&
\ph^{(1)} (k)
= \frac{\ka^4 M}{4} \Big( \frac43 \beta_2 - \frac13 \beta_0
- \beta_{\scriptscriptstyle RT1} - \beta_{\scriptscriptstyle RT2} \Big)
\ln \Big( \frac{k^2}{\mu^2} \Big),
\nn
\\
&&
\psi^{(1)} (k)
=  \frac{\ka^4 M}{4}
\Big( \frac23 \beta_2 + \frac13 \beta_0 - \beta_{\scriptscriptstyle RT1} + \beta_{\scriptscriptstyle RT2} \Big)
\ln \Big( \frac{k^2}{\mu^2} \Big),
\eeq
where $\,k = \vert \vec{k} \vert $. Next, using the following inverse Fourier transforms,
\beq
&&
\int \frac{d^3 k}{(2\pi)^3}
e^{-i \vec{k} \cdot \vec{r}} \frac{1}{k^2}
= \frac{1}{4\pi r},
\nn
\\
\label{log}
&&
\int \frac{d^3 k}{(2\pi)^3} \,
e^{-i \vec{k} \cdot \vec{r}} \ln \Big( \frac{k^2}{\mu^2} \Big)
= - \frac{1}{2\pi r^3}
\quad (r \neq 0) \, ,
\eeq
we arrive at the general results
\beq
&& \label{phi}
\ph(r)\, = \,  - \frac{\ka^2 M}{16 \pi r}
\,-\, \Big(\frac43 \beta_2 - \frac13 \beta_0
- \beta_{\scriptscriptstyle RT1} - \beta_{\scriptscriptstyle RT2} \Big) \frac{\ka^4 M}{8 \pi r^3} \, ,
\\
&&
\label{psi}
\psi(r) =  - \frac{\ka^2 M}{16 \pi r}
-  \Big(\frac23 \beta_2 + \frac13 \beta_0 - \beta_{\scriptscriptstyle RT1}
+ \beta_{\scriptscriptstyle RT2} \Big) \frac{\ka^4 M}{8 \pi r^3}.
\eeq

We do not write the explicit expression for the potentials
in terms of the parameters $\ga_{1,\ldots,6}$, $\ga_0$ and $\la$
since they are too cumbersome and not very illuminating.
At the same time, it is easy to see, using Eqs.~\eq{beta1}--\eq{beta5},
that the obtained potentials depend on the parametrization and gauge-fixing
parameters. On the other hand, the potentials $\ph(r)$ and $\psi(r)$
do not depend on the parametrization of the background field
\eq{background}. Indeed, the dependence on the parametrization
of the quantum fields and the gauge-fixing should be expected,
because the effective equations of motion are gauge and
parametrization dependent~\cite{Vilk-Uni}.

In the standard $g_{\mu\nu} + \phi_{\mu\nu}$ parametrization limit,
$\ga_1 \to 1$, $\ga_{2,\ldots,6} \to 0$ and $\ga_0 \to 0$. Then,
Eqs.~\eq{phi} and \eq{psi} become
\begin{equation}
\label{phi-standard}
\ph(r) =  - \frac{G M}{r} \left( 1 + \frac{61}{60}\,\frac{G}{\pi r^2}
\right),
\end{equation}
\begin{equation}
\psi(r) =  - \frac{G M}{r} \left( 1 +
\frac{23}{60}\,\frac{G}{\pi r^2}
\right).
\end{equation}
Formula \eq{phi-standard} matches
the result presented in~\cite{DM97}
for the $h_{00}$ component of the
background metric evaluated
in the standard parametrization with
the minimal DeWitt gauge.

\section{The motion of a test particle}
\label{sec4}

Let us now follow the main idea of Ref.~\cite{DM97} that even though
the loop corrections to the gravitational field generated by a
point-like mass is parametrization and gauge-fixing dependent, the
physical observables should be invariant. As an example, consider
the acceleration of a test particle moving in the gravitational field
described in the previous section. The key observation is that the
test particle also couples with the quantum metric and, as a
consequence of this, its geodesic equation receives quantum
corrections. Thus, the two types of quantum corrections are supposed
to combine into the invariant result. Finally, in the non-relativistic
limit, we expect to meet invariant quantum corrections to
Newton's law, $m \vec{a} = \vec{F}$.

The classical action for a test particle is given by equation (\ref{actM}), but with
another mass,
\beq
\label{action_test}
S_m \,=\, - \,m \, \int \sqrt{g_{\mu\nu} \, d x^\mu d x^\nu},
\eeq
and coordinates $x^\mu = (t, \vec{r})$. We assume that
$m \ll M$, such that we can neglect contribution of the small mass
$m$ to the potentials \eq{phi} and \eq{psi}. The one-loop corrections
to \eq{action_test} can be obtained from \eq{EA-Weyl} by the
substitution
\ $T^{\mu\nu} \longrightarrow T^{\mu\nu} + T^{\mu\nu}_m$~\cite{DM97},
where
\beq
T^{\mu\nu}_m  = m \int ds \,
\de \left(y-x(s) \right) u^\mu u^\nu.
\eeq
Applying this procedure yields
\beq
\label{EA_test}
\bar{\Ga}^{(1)}_m = \int d^4 x \sqrt{-g} \,
\Big\{
\ka^2 \beta_{\scriptscriptstyle RT1}  R_{\mu\nu} \ln \Big( \frac{\Box}{\mu^2} \Big) T^{\mu\nu}_m
- \frac12 \ka^2 \beta_{\scriptscriptstyle RT2}  R \ln \Big( \frac{\Box}{\mu^2} \Big) T_m
- \ka^4 \beta_{\scriptscriptstyle TT} T \ln \Big( \frac{\Box}{\mu^2} \Big) T_m
\Big\}.
\eeq

The total action for the test particle is
$\Ga_m = S_m + \bar{\Ga}^{(1)}_m$.
Taking the functional derivative with respect to $x_\mu$, we find
\beq
\frac{1}{m} \,\frac{\de \Ga_m}{\de x_\mu}
\,=\, - \, \left( \frac{d^2 x^\mu}{ds^2}
+ \Ga^{\mu}_{\al\be}
\frac{d x^\al}{ds} \frac{d x^\be}{ds} \right)
+ \frac{1}{m} \frac{\de \bar{\Ga}^{(1)}_m}{\de x_\mu}
\,=\, 0,
\eeq
where $\Ga^{\mu}_{\al\be}$ is the Christoffel symbol. In the
non-relativistic limit, this equation boils down to
\beq
\label{quantum_Newton}
\vec{a} - \frac{1}{m} \frac{\de \bar{\Ga}^{(1)}_m}{\de \vec{r}}
= - \vec{\na} \ph,
\qquad
\mbox{where}
\qquad
\vec{a} = \frac{d^2 \vec{r}}{d t^2}
\eeq
is the particle acceleration.

To evaluate the functional derivative in~\eq{quantum_Newton},
all the metric functions in Eq.~\eq{EA_test} should be taken at the
order ${\mathcal O}(\hbar^0)$, owing to the one-loop approximation
and the fact $\beta_i \sim {\mathcal O}(\hbar)$. On top of that, in the
non-relativistic limit we can take $s = t$, $u^\mu = (1,0)$.
Thus, using Eq.~\eq{EMT_static} for the source energy-momentum tensor, we get for~\eq{EA_test}, in the weak-field
approximation,
\beq
\bar{\Ga}^{(1)}_m
&=&
m \int dt \, \Big\{
\ka^2 \beta_{\scriptscriptstyle RT1} \ln \Big( \frac{-\De}{\mu^2} \Big)
R_{00}^{(0)} \left(\vec{r}(t) \right)
- \frac12 \ka^2 \beta_{\scriptscriptstyle RT2} \ln \Big( \frac{-\De}{\mu^2} \Big)
R^{(0)} \left(\vec{r}(t) \right)
\nn
\\
&&
\quad
- \,\,
\ka^4 \beta_{\scriptscriptstyle TT} M \ln \Big( \frac{-\De}{\mu^2} \Big)
\de \left(\vec{r}(t) \right) \Big\}.
\eeq
The ${\mathcal O}(\hbar^0)$ curvatures can be evaluated
through the relations for the $00$-component of the Ricci
tensor and for the scalar curvature and also using Eq.~\eq{0-order},
\begin{align}
&
R_{00}^{(0)} \left(\vec{r}(t) \right) =
\De \ph^{(0)} = \frac{\ka^2 M}{4} \de \left(\vec{r}(t) \right),
\nn
\\
&
R^{(0)} \left(\vec{r}(t) \right) = 2 ( \De \ph^{(0)}   - 2 \De \psi^{(0)} )
= - \frac{\ka^2 M}{2} \de \left(\vec{r}(t)\right).
\end{align}

Finally, using the Eq.~\eq{log} in the coordinate space,
\begin{equation}
\ln \Big( - \frac{\De}{\mu^2}  \Big) \de  (\vec{r})
\,=\, - \,\frac{1}{2 \pi r^3},
\end{equation}
we find
\begin{equation}
\begin{split}
\bar{\Ga}^{(1)}_m = - \frac{\ka^4 m M}{8 \pi}
(-\beta_{\scriptscriptstyle RT1} - \beta_{\scriptscriptstyle RT2} + 4 \beta_{\scriptscriptstyle TT} )
\int \frac{dt}{r^3(t)},
\end{split}
\end{equation}
implying that
\begin{equation}
\frac{1}{m} \frac{\de \bar{\Ga}^{(1)}_m}{\de \vec{r}}  =
-\frac{\ka^4 M}{8 \pi} (-\beta_{\scriptscriptstyle RT1} - \beta_{\scriptscriptstyle RT2} + 4 \beta_{\scriptscriptstyle TT})
\vec{\na} \left(
\frac{1}{r^3}
\right).
\end{equation}
Plugging this equation into~\eq{quantum_Newton} and using
Eq.~\eq{phi}, we get
\begin{equation}
\label{ace_final}
\vec{a} = - \vec{\na} \left[- \frac{\ka^2 M}{16 \pi r}
-
\Big(\frac43 \beta_2 - \frac13 \beta_0 - 2\beta_{\scriptscriptstyle RT1}
- 2\beta_{\scriptscriptstyle RT2} + 4 \beta_{\scriptscriptstyle TT} \Big) \frac{\ka^4 M}{8 \pi r^3 } \right].
\end{equation}

The quantum corrected Newtonian potential is defined through
the relation~\cite{DM97}
\begin{equation}
\label{Newton_definition}
\vec{a} \equiv  - \vec{\na} V.
\end{equation}
Thus, \eq{ace_final} and \eq{Newton_definition} give
\beq
V(r) = - \frac{\ka^2 M}{16 \pi r}
-
\Big(\frac43 \beta_2 - \frac13 \beta_0
- 2 \beta_{\scriptscriptstyle RT1} - 2\beta_{\scriptscriptstyle RT2} + 4 \beta_{\scriptscriptstyle TT} \Big)
\frac{\ka^4 M}{8 \pi r^3 } \, .
\label{potential_final}
\eeq
Finally, using Eqs.~\eq{beta1}--\eq{beta5}
we get
\beq
V(r)\, =\,
- \frac{G M}{r} \left( 1 +
\frac{17}{20}\,\frac{G}{\pi r^2}
\right),
\label{NewNew}
\eeq
which does not depend on parametrization or gauge parameters.
Let us note that the cancellation of the gauge and parametrization
dependencies in this expression looks rather impressive, taking
the (somehow scaring) form of the expressions in Appendix A.

The coefficient $17/20$ does not agree with the value presented
in Ref.~\cite{DM97}. This discordance is due to a missing factor
of four in formula (35) of~\cite{DM97}, as we explain in the
Appendix B. It is remarkable that the complicated calculation in
this work has only this small mistake and is otherwise correct.

\section{General proof of gauge independence}
\label{sec5}

We have evaluated the coefficients $\beta_i$ explicitly in the most
general parametrization for the quantum field and the particular
case of massive point-like masses. However, given Eq.~\eq{div_Weyl},
we provided an expression for  the Newtonian potential \eq{potential_final}
independently on
the explicit values of the $\beta_i$. Therefore,
to prove that the potential~\eq{potential_final} does not depend
on the gauge-fixing and parametrization choices, it is enough to
prove that the following combination,
\beq
\label{betainv}
\beta_{\text{inv}}
=
\frac43 \beta_2 - \frac13 \beta_0 - 2\beta_{\scriptscriptstyle RT1} - 2\beta_{\scriptscriptstyle RT2} + 4 \beta_{\scriptscriptstyle TT} \, ,
\eeq
is a gauge and parametrization invariant quantity. For this, we follow
\cite{frts82,a} and \cite{JDG-QG}, i.e., we employ the general statement
about the gauge-fixing and parametrization independence of the
on-shell effective action. In particular, the difference between the
divergences of two versions of the one-loop effective action,
evaluated using different gauge and parametrizations $\alpha_i$ and $\alpha_0$,
is proportional to the classical equations of motion,
\beq
&&
\de \bar{\Ga}_{div}^{(1)} =
\bar{\Ga}_{div}^{(1)} (\alpha_i) - \bar{\Ga}_{div}^{(1)} (\alpha_0)
= - \frac{\mu^{n-4}}{(4 \pi)^2 (n-4)}  \int d^n x \sqrt{-g} \, \vp^{\mu\nu} f_{\mu\nu},
\nn
\\
&&
\vp^{\mu\nu}
= R^{\mu\nu} - \frac12 g^{\mu\nu} R - \frac{\ka^2}{2} T^{\mu\nu}.
\label{deGamma}
\eeq

As the divergences are local and covariant quantities with mass dimension four, the tensor function $f_{\mu\nu}$ has the
following general structure
\begin{equation}
\label{fmn}
f_{\mu\nu} = b_1 R_{\mu\nu}
+ b_2 R  g_{\mu\nu}
+ \ka^2 b_3 T_{\mu\nu}
+ \ka^2 b_4 T g_{\mu\nu},
\end{equation}
where the parameters $b_{1,2,3,4}$ depend on the choice of the
gauge and parametrization parameters~$\al_i$. Thus, replacing
\eq{fmn} in \eq{deGamma} and using the definition~\eq{part_T_square}
we get:
\begin{equation}
\begin{split}
\de \bar{\Ga}_{div}^{(1)} = & \,\,
- \frac{\mu^{n-4}}{(4 \pi)^2 (n-4)} \int d^n x \sqrt{-g}\,
\Big\{
b_1 R_{\mu\nu}^2
- \big(\tfrac12 b_1 + b_2\big) R^2
+  \ka^2 \big(b_3 - \tfrac12 b_1\big) R_{\mu\nu} T^{\mu\nu}
\\
&
\quad
- \,\,\ka^2 \big(\tfrac12 b_2 + \tfrac12 b_3 + b_4\big) R T
- \ka^4\big(\tfrac12 b_3 + \tfrac12 b_4\big)  T^2
\Big\}.
\end{split}
\end{equation}
So, under a gauge and/or parametrization transformation, the
coefficients in \eq{divs-1} transform according to:
\beq
&&
c_1 \to c_1,
\quad
c_2 \to c_2 + b_1,
\quad
c_3 \to c_3 - \left(\frac{b_1}{2} + b_2\right) \, ,
\quad
c_4 \to c_4 + \left(b_3 - \frac{b_1}{2}\right) \, ,
\nn
\\
&&
c_5 \to c_5 - \frac{1}{2} ( b_2 + b_3 + 2 b_4) \, ,
\quad
c_6 \to c_6 - \frac{1}{2} (b_3 + b_4) \, .
\eeq
It is a simple exercise to check that, besides $c_1$, there is the
following gauge- and parametrization-independent combination:
\beq
\label{cinv}
c_{\text{inv}} = c_2+ c_3 + c_4 - 2 c_5 + 4 c_6.
\eeq
In particular, in the notations of \eq{div_Weyl}, the formula
\eq{cinv} implies that \eq{betainv} is truly gauge and
parametrization invariant.
This also means the physical results derived on the
basis of usual effective action and for the unique Vilkovisky-DeWitt
effective action \cite{Vilk-Uni} will be the same. The reader can
consult \cite{bavi85} for the first calculation in quantum gravity in this formalism and the more recent Ref.~\cite{BTS-Vil-1} for the explicit verification of the parameterization independence,
including the conformal mode (\ref{gen-bfm}).

In fact, the generalization of the calculations done in~\cite{BTS-Vil-1}
to include the massive point particles is straightforward.
Without entering in the full details, the result for the one-loop divergences
in the theory \eq{classical_action}, using the Vilkovisky-DeWitt definition
for the effective action and the general parametrization \eq{gen-bfm}, has the
same form as Eq.~\eq{divs-1}, but now with the following gauge and parametrization independent
coefficients,
\begin{align}
c_1 = \frac{53}{45},
\qquad
c_2 = -\frac{61}{90},
\qquad
c_3 = \frac{25}{36},
\qquad
c_4 =  -\frac{3}{2},
\qquad
c_5 = \frac{1}{2},
\qquad
c_6 = - \frac{5}{32}.
\label{c123456-Vil}
\end{align}
Using (\ref{beta5}) and (\ref{potential_final}), it is easy
to verify that \eq{c123456-Vil} provides the same Newtonian
potential of Eq.~\eq{NewNew}.
Thus, the cancellation observed in
(\ref{NewNew}) is a necessary feature of effective quantum
gravity, especially in the case when only metric is a quantum
field, while the macroscopic massive bodies are treated as
classical sources, according to the arguments of \cite{DM97}
and \cite{Polemic}.

As the massive sources are regarded classical, one can certainly
expect the parametrization independence to hold for other
configurations of matter, including non-point mass distributions,
such as, e.g., dark matter.

\section{Universality of the effective field theory result}
\label{secLM}

The effective field theory approach cannot be taken lightly, as we cannot write down all possible action operators without having any guiding principle. Indeed, it is well known that any higher derivative theory has a number of ghost-like states depending on the number of derivatives that catastrophically make the vacuum decay instantaneously in ghost and normal particles. This is a trivial consequence of the energy conservation since ghosts carry negative energy~\cite{Cline:2003gs}.
Therefore, EFTs make sense and are well defined only if they turn out to be the low-energy approximation, or large distance limit, of more fundamental theories consistent with unitarity (perturbative), and (super-)renormalizability or finiteness.

One example of finite and unitarity theory is string theory, but we here focus on ultraviolet complete gravitational theories in the quantum field theory framework. In the latter case, we have two classes of well-defined theories: higher derivative theories {\em \`a la} Lee-Wick~\cite{LW1, LW2, ShapiroModestoLW, ModestoLW, Anselmi:2017ygm} or more general higher derivative theories \cite{shapiro3}, in which unitarity can be formally achieved by means of the
Anselmi-Piva prescription, which is an extension of conventional
Wick rotation, operational for the higher derivative models
\cite{AnselmiPiva1,AnselmiPiva2,AnselmiPiva3}; or nonlocal theories without ghost-like perturbative degrees of freedom \cite{kuzmin,Modesto, ModestoLeslawF,ModestoLeslawR}. In Lee-Wick quantum gravity, the ghost particles may only have complex mass square~\cite{ShapiroModestoLW, ModestoLW}, while in general higher derivative theories \cite{shapiro3} we also have real ghosts, but in both cases, such states do not appear as asymptotic states. Indeed, such homogeneous solutions of the linear equations of motion can be removed from the spectrum of the theory by hand and, most importantly, they are not created again in the loop amplitudes once the Anselmi-Piva prescription is implemented. This is very similar to what happens with the BRST ghosts that appear because of the quantum gauge invariance. Indeed, we can safely fix the number of external BRST ghosts to zero, because in the loop amplitudes, they are exactly canceled by the non-physical polarization states of the gauge bosons. Similarly, in higher derivative theories, we can fix the number of external ghosts to zero because they are not regenerated in the loop amplitudes at any perturbative order whether the prescription~\cite{AnselmiPiva1,AnselmiPiva2,AnselmiPiva3} is implemented.

Differently, in nonlocal quantum gravity \cite{kuzmin, Modesto, ModestoLeslawF,ModestoLeslawR}, the perturbative degrees of freedom are the same as the local theory \cite{Briscese:2018bny,Briscese:2019rii}
and the convergence of the loop amplitudes is achieved by introducing an exponential (but asymptotically polynomial \cite{kuzmin, Modesto}) form factor that has no poles in the whole complex plane. At the quantum level, the loop amplitudes are computed in the Euclidean signature, and afterwards, the physical amplitudes are obtained using an analytic continuation of the external energies from the Euclidean space to Minkowski spacetime \cite{Briscese:2018oyx, Briscese:2021mob}.
A unified nonlocal theory of all fundamental interactions with the aforementioned properties of the purely gravitational theories has been recently proposed in \cite{Modesto:2021okr, Modesto:2021ief}. In the ultraviolet regime all nonlocal gravitational theories (or local with more than ten derivatives in four dimensions) are asymptotically free \cite{Briscese:2019twl} or finite \cite{ModestoLeslawF,Rachwal:2021bgb}.
Theories with a number of derivatives from six to eight are super-renormalizable but, to prove asymptotic freedom, quantitative analysis and explicit computations are needed because there are divergences also at two and/or three loops.
The last theory which deserves special attention is the four derivatives
or Stelle's theory \cite{Stelle}.

Said this, it is very interesting to investigate the infrared properties of any of the theories described above. In particular, in this work, we are interested in the leading correction to the Newtonian potential. Such correction is strictly related to the analytic properties of the one-loop amplitude regardless of the renormalizability or finiteness of the theory. Indeed, the analyticity only depends on the Landau singularities of the one-loop amplitude, namely when the denominators of the amplitude are zero for the same value of the external energy. This property depends only on the low-energy spectrum of the theory, and, hence, the contribution of the massive states is sub-dominant. Therefore, the main corrections to the quantum effective action at large distance come only from the massless states.

As an example we can consider a nonlocal scalar theory with an exponential form factor~\cite{spallucci}.
This theory is ultraviolet finite, but the one-loop amplitude at low energy, namely when $p^2/\Lambda^2 \lesssim 1$ (here $\Lambda$ is the non-locality scale), shows up exactly the same logarithmic non-analyticity of the local theory. This result is understandable whether we identify $\Lambda$ with the cut-off of the local theory. 

Let us expand on the case of the scalar model. The nonlocal Lagrangian that we would like to consider reads:
\beq
\mathcal{L}_{\phi} = -\frac{1}{2}  \phi \, {\rm e}^{H(- \Box/\Lambda^2)}\left(\Box +  m^2 \right)\phi  - \frac{\lambda}{3 !}  \phi^3 \, ,
\label{phin1}
\eeq
where $H(z)$ is an analytic function without poles in the complex plane at finite distance. The propagator of the scalar field is:
\beq
G(k) = \frac{{\rm i}\,  {\rm e}^{-H(k)}  }{k^2 - m^2 + {\rm i} \epsilon } \, .
\label{propag}
\eeq
The one-loop amplitude turns out to be:
\beq
{\mathcal M} = \frac{\lambda^2}{32 \pi^4} \int_{-\infty}^{+ \infty} d^4 k_{\rm E} \,
 \frac{  {\rm e}^{- H(k_{\rm E}) } \, {\rm e}^{- H(k_{\rm E} - p_{\rm E}) } }{ (k_{\rm E}^2+m^2) \, [(k_{\rm E} + p_{\rm E} )^2 +m^2 ]} \, ,
 \label{1loopG}
\eeq
where $k_{\rm E}$ and $p_{\rm E}$ are respectively the internal and external purely imaginary four-momenta.
For the simplest non-locality, $H =  \Box = - k^2/\Lambda^2 = k_{\rm E}^2/\Lambda^2$, the above amplitude (\ref{1loopG}) turns into a simple Gaussian integral. The explicit computation was done for the first time in \cite{spallucci} and more recently in \cite{liqiang}. The result after the analytic continuation from purely imaginary to real external energy reads:
\beq
\mathcal{M} = - \frac{\lambda^2}{16 \pi^2} \int_0^{1/2} d x \, {\rm Ei} \left[ - \left(\frac{ m^2  - p^2  x (1-x)}{x \Lambda^2} \right)\right] \, ,
\label{Mexact}
\eeq
where ${\rm Ei}(z)$ is the exponential integral.
Notice that the result is finite without need of any regularization.

We now consider the low energy limit $p^2/\Lambda^2 \lesssim 1$. In order to investigate such limit, we remind the Taylor expansion of the exponential integral ${\rm Ei}(z)$,
\beq
{\rm Ei}(z) = \gamma_{\rm E} + \ln z + z + \frac{z^2}{4} + \dots \, .
\eeq
Using the above expansion in (\ref{Mexact}) we get:
\beq
\mathcal{M} \simeq - \frac{\lambda^2}{32 \pi^2} \int_0^{1} d x \, {\ln}  \left(\frac{ m^2  - p^2  x (1-x)}{ \Lambda^2} \right) \, .
\label{Mexpan}
\eeq
Therefore, ({\ref{Mexact}) at low energy resembles the result in the local quantum field theory up to proper identification of the non-locality scale with the cut-off of the local renormalizable theory, which we give here for completeness:
\beq
\Lambda^2 = \Lambda_{\rm cut-off} \, {\rm e}^{\gamma_{\rm E} + 1 + \ln 2} \, .
\eeq

As another example, let us now consider a nonlocal theory without perturbative degrees of freedom around the usual trivial background
$\phi_{\rm B} = 0$. The theory reads:
\beq
\mathcal{L}_{\phi} = -\frac{1}{2}  \phi \, {\rm e}^{H(- \Box/\Lambda^2)}  m^2 \, \phi  - \frac{\lambda}{3 !}  \phi^3 \, .
\label{phinTOP}
\eeq
Clearly, at the zero-order in $\lambda$ the perturbative solution of the free-theory equation of motion is $\phi = 0$, namely
\beq
{\rm e}^{H(- \Box/\Lambda^2)}  \, m^2 \, \phi = 0 \quad \Longrightarrow \quad \phi = 0 .
\label{classic}
\eeq
Therefore, there are no perturbative degrees of freedom propagating around the trivial background.
Equivalently, the propagator in momentum space, which can be obtained from (\ref{propag}) removing $k^2$ from the denominator, has no poles in the whole complex plane at finite energy,
\beq
G(k)  = - \frac{\rm i}{m^2} {\rm e}^{-H(k^2/\Lambda^2)} \, ,
\eeq
consistently with the classical solution in (\ref{classic}).
Now we are ready to compute one-loop bubble diagram for the theory (\ref{phinTOP}). The one-loop amplitude reads:
\beq
{\mathcal M} =  - \frac{\lambda^2}{32 \pi^4} \int_{-\infty}^{+ \infty} d^4 k_{\rm E} \,
 \frac{ {\rm i} \, {\rm e}^{- H(k_{\rm E}) }}{ m^2 } \, \frac{ {\rm i} \, {\rm e}^{- H(k_{\rm E} - p_{\rm E}) } }{ m^2 } \, .
 \label{Ampli}
\eeq
%
Again for the case of a Gaussian non-locality, namely $H =  \Box + m^2 = (- k^2+m^2)/\Lambda^2 = (k_{\rm E}^2+m^2)/\Lambda^2$, the amplitude (\ref{Ampli}) simplifies to:
\beq
{\mathcal M} & = & - \frac{\lambda^2}{32 \pi^4} \int_{-\infty}^{+ \infty} d^4 k_{\rm E} \,
 \frac{ {\rm i} \, {\rm e}^{- \frac{k_{\rm E}^2+m^2}{\Lambda^2} }}{ m^2 }
 \, \frac{ {\rm i} \,  {\rm e}^{- \frac{( k_{\rm E}-p_{\rm E})^2+m^2}{\Lambda^2} }  }{ m^2 } \nonumber\\
 & = & \frac{\lambda^2}{128 \pi^2} {\rm e}^{ - \frac{1}{2 \Lambda^2} \left( - p^2 + (2 m)^2 \right) } \, ,
 \label{AmpliG}
\eeq
which is analytic in the whole complex plane according to the absence of perturbative degrees of freedom in the theory (\ref{phinTOP}).

We conclude that the logarithm correction to the quantum effective action is universally related to the low energy spectrum of the theory regardless of the ultraviolet completion of the theory.

The same argument applies to a general nonlocal or higher derivative theory, being finite or (super-)renormalizable, with the graviton the only massless state in the spectrum. The action for a general local higher derivative gravitational theory reads:
\beq
S_{\rm HD} = \int d^D x \sqrt{ - g}  \left[ \kappa^{-2} R
+ \sum_{n=0}^{\rm N} \left(\omega_{0, n} R \, \Box^n R
+ \omega_{2, n} R_{\mu\nu} \, \Box^n R^{\mu\nu} \right)
+ V\left( {\mathcal O}\big({\rm Riem}^3 \big) \right)
\right]  ,
\label{HD}
\eeq
where the last term plays the role of a potential because it is at least
cubic in the Riemann tensor and, then, it does not contribute to the
propagator around the Minkowski's background. Moreover, the potential
contains at most $2 {\rm N} + 4$ derivatives in order to not spoil
the super-renormalizability of the theory. However, the presence of
$V\left( {\mathcal O}\big({\rm Riem}^3 \big) \right)$ is crucial to
making the theory finite \cite{Modesto,ModestoLeslawF}.
The particle spectrum of the higher derivative action (\ref{HD})
contains a massless graviton and a finite number of massive states
(including complex and real ghosts) whose masses are defined in
terms of the 
constants $\kappa$, $\omega_{0,n}$, and $\omega_{2,n}$.
As said above, the theory (\ref{HD}) can be finite, namely all
the loop amplitudes are convergent. However, at an energy scale
much smaller than the mass of all the massive states, the one-loop
quantum effective action has the same form as (\ref{EA-Weyl})
for a proper identification (which means up to an overall
dimensionless rescaling) of the renormalization group invariant
scale $\mu$ in (\ref{EA-Weyl}) with the mass scale that is
implicitly hidden in the parameters $\omega_{0,n}$ and
$\omega_{2,n}$. In other words, at low energy,
the higher derivative operators play the role of a higher
derivative regularization. Once again, this is not surprising
because they are related to the analytic properties of the one-loop
quantum effective action that only depends on the Landau
singularities of the amplitude. Moreover, in the low energy
limit the contribution of the massive states is sub-leading
with respect to the massless graviton.

\section{Conclusions and discussions}
\label{sec6}

We have reconsidered the arguments for taking into account only the
gravitational loops in effective quantum gravity. The physical reasons
and, in particular, the analogy with the Casimir force show that the
massive sources should not be quantized. Taking this into account,
it is well-known that the numerical effects of quantum corrections
are too small to be measured in the laboratory. However, this does not
make irrelevant the consistent derivation of the quantum corrections
and the discussion of their (non)universality.

The main part of the work reports on a calculation similar to the
previously done in \cite{DM97}, but in an arbitrary parametrization of
quantum perturbation of the metric. As a result, we have found that the effective
quantum gravity correction to Newton's gravitational potential
is completely independent of the choice of the parametrization. Together
with the mentioned qualitative arguments and the gauge-fixing
independence established in \cite{DM97}, our result ensures that
the final outcome (\ref{potential_final}) is correct. In principle, this
result can be generalized for more complicated distributions of
mass. In this respect, the more general proof of invariance given
in Sec.~\ref{sec5} is relevant, as it provides a good expectation
to obtain sound results.

Furthermore, we discussed in Sec.~\ref{secLM} the universality and independence of the leading correction to Newton's potential in the effective field theories resulting in the low-energy limit of ultraviolet complete theories of quantum gravity in the quantum field theory framework. In particular, we showed that the leading quantum correction to Newton's potential is strictly related to the low-energy spectrum of the fundamental theory rather than the details of the ultraviolet complete quantum gravity theory, in accordance with the general effective quantum field theory approach.
Indeed, the logarithm quantum correction is related to the Landau singularities and the analytic properties of the one-loop amplitude, which only depend on the low-energy spectrum of the classical theory. Therefore, not all the quantum effective theories produce the same result. The latter statement has been shown explicitly with simple examples.

Taking the parametrization and gauge invariance, the direct relation to
the beta functions, and the fact that the quantum
corrections are the same as in the usual and Vilkovisky-DeWitt
versions of the effective actions, the main result for the quantum
corrected potential can be compared to the invariant version of the
renormalization group improved classical action \cite{TV90,UEA-RG}.
In the present framework, when dealing with the Newtonian interaction
between two massive particles, the interpretation of the
renormalization group scale is straightforward. This interpretation
is much more subtle and complicated in other physical situations,
e.g., in Cosmology \cite{RotCurves}, where some specially
designed procedures can be applied \cite{HS}, but still do not
guarantee unique interpretation of the scale. Thus, it would be
interesting to compare the two ways of scale identification in the
cases like the one considered above, with a well-defined
procedure \cite{DM97}.

In conclusion, let us note that the main approximation in effective
quantum gravity is supposed to hold between the Planck scale
in the UV and the Hubble scale in the IR . In particular, this
interval includes all the physically and astronomically interesting
high-energy gravitational phenomena related to early cosmology
and black hole physics. Thus, there still is a possibility to find
applications to the results based on correctly defined quantum
gravity corrections to classical gravity.

\section*{Acknowledgements}
\label{secAck}

Authors are grateful to Michael Bordag, Michael Duff and Diego
Mazzitelli for useful and friendly discussions.
I.Sh. was partially supported by Conselho Nacional de
Desenvolvimento Cient\'{i}fico e Tecnol\'{o}gico - CNPq (Brazil)
under the grant 303635/2018-5, by Funda\c{c}\~{a}o de Amparo \`a Pesquisa
de Minas Gerais - FAPEMIG under the project PPM-00604-18, and by
the Ministry of Education of Russian Federation
under the project No. FEWF-2020-0003.
L.M. was supported by the Basic Research Program of Science,
Technology and Innovation Commission of Shenzhen Municipality
(grant No. JCYJ20180302174206969).

\newpage
\section*{Appendix A}
\label{ApA}

The explicit expressions for the coefficients in \eq{divs-1} were obtained using \textsc{Wolfram Mathematica}~\cite{Mathematica} and the auxiliary tensor algebra package \textsc{xAct}~\cite{xAct,xTras}.
For the sake of completeness, we present the list of the coefficients
used in Eq.~(\ref{c123456}):
\begin{equation}
    \begin{split}
    \xi_2 = & \,\,
    16 \lambda^2 \gamma_0^2 \gamma_{1}^4 + 8 \lambda \gamma_0 \gamma_{1}^3 \gamma_{3} - 128 \lambda^2 \gamma_0^2 \gamma_{3}^2 - 32 \lambda \gamma_0 \gamma_{1} \gamma_{3}^2 -  \gamma_{1}^2 \gamma_{3}^2 - 128 \lambda \gamma_0 \gamma_{2} \gamma_{3}^2 - 16 \gamma_{1} \gamma_{2} \gamma_{3}^2  \\
    & - 32 \gamma_{2}^2 \gamma_{3}^2 + 16 \lambda \gamma_0 \gamma_{1}^3 \gamma_{5} + 4 \gamma_{1}^2 \gamma_{3} \gamma_{5} + 4 \gamma_{1}^2 \gamma_{5}^2
    ,
    \end{split}
\end{equation}
\begin{equation}
    \begin{split}
    \xi_3 = & \,\,
    3584 \lambda^4 \gamma_0^4 \gamma_{1}^4 + 1024 \lambda^3 \gamma_0^3 \gamma_{1}^5 + 64 \lambda^2 \gamma_0^2 \gamma_{1}^6 - 4 \lambda \gamma_0 \gamma_{1}^7 + 4096 \lambda^3 \gamma_0^3 \gamma_{1}^4 \gamma_{2} + 512 \lambda^2 \gamma_0^2 \gamma_{1}^5 \gamma_{2}  \\
    & - 48 \lambda \gamma_0 \gamma_{1}^6 \gamma_{2} + 1024 \lambda^2 \gamma_0^2 \gamma_{1}^4 \gamma_{2}^2 - 192 \lambda \gamma_0 \gamma_{1}^5 \gamma_{2}^2 - 256 \lambda \gamma_0 \gamma_{1}^4 \gamma_{2}^3 - 55296 \lambda^4 \gamma_0^4 \gamma_{1}^2 \gamma_{3}  \\
    & - 26112 \lambda^3 \gamma_0^3 \gamma_{1}^3 \gamma_{3} - 4736 \lambda^2 \gamma_0^2 \gamma_{1}^4 \gamma_{3} - 392 \lambda \gamma_0 \gamma_{1}^5 \gamma_{3} - 14 \gamma_{1}^6 \gamma_{3} - 110592 \lambda^3 \gamma_0^3 \gamma_{1}^2 \gamma_{2} \gamma_{3}  \\
    & - 39936 \lambda^2 \gamma_0^2 \gamma_{1}^3 \gamma_{2} \gamma_{3} - 4928 \lambda \gamma_0 \gamma_{1}^4 \gamma_{2} \gamma_{3} - 220 \gamma_{1}^5 \gamma_{2} \gamma_{3} - 82944 \lambda^2 \gamma_0^2 \gamma_{1}^2 \gamma_{2}^2 \gamma_{3}  \\
    & - 20352 \lambda \gamma_0 \gamma_{1}^3 \gamma_{2}^2 \gamma_{3} - 1304 \gamma_{1}^4 \gamma_{2}^2 \gamma_{3} - 27648 \lambda \gamma_0 \gamma_{1}^2 \gamma_{2}^3 \gamma_{3} - 3456 \gamma_{1}^3 \gamma_{2}^3 \gamma_{3}  \\
    & - 3456 \gamma_{1}^2 \gamma_{2}^4 \gamma_{3} + 30720 \lambda^4 \gamma_0^4 \gamma_{3}^2 + 15360 \lambda^3 \gamma_0^3 \gamma_{1} \gamma_{3}^2 + 3072 \lambda^2 \gamma_0^2 \gamma_{1}^2 \gamma_{3}^2 + 288 \lambda \gamma_0 \gamma_{1}^3 \gamma_{3}^2  \\
    & + 12 \gamma_{1}^4 \gamma_{3}^2 + 61440 \lambda^3 \gamma_0^3 \gamma_{2} \gamma_{3}^2 + 23040 \lambda^2 \gamma_0^2 \gamma_{1} \gamma_{2} \gamma_{3}^2 + 3072 \lambda \gamma_0 \gamma_{1}^2 \gamma_{2} \gamma_{3}^2 + 144 \gamma_{1}^3 \gamma_{2} \gamma_{3}^2  \\
    & + 46080 \lambda^2 \gamma_0^2 \gamma_{2}^2 \gamma_{3}^2 + 11520 \lambda \gamma_0 \gamma_{1} \gamma_{2}^2 \gamma_{3}^2 + 768 \gamma_{1}^2 \gamma_{2}^2 \gamma_{3}^2 + 15360 \lambda \gamma_0 \gamma_{2}^3 \gamma_{3}^2 + 1920 \gamma_{1} \gamma_{2}^3 \gamma_{3}^2  \\
    & + 1920 \gamma_{2}^4 \gamma_{3}^2 - 221184 \lambda^4 \gamma_0^4 \gamma_{1}^2 \gamma_{4} - 110592 \lambda^3 \gamma_0^3 \gamma_{1}^3 \gamma_{4} - 20480 \lambda^2 \gamma_0^2 \gamma_{1}^4 \gamma_{4} - 1664 \lambda \gamma_0 \gamma_{1}^5 \gamma_{4}  \\
    & - 56 \gamma_{1}^6 \gamma_{4} - 442368 \lambda^3 \gamma_0^3 \gamma_{1}^2 \gamma_{2} \gamma_{4} - 165888 \lambda^2 \gamma_0^2 \gamma_{1}^3 \gamma_{2} \gamma_{4} - 20480 \lambda \gamma_0 \gamma_{1}^4 \gamma_{2} \gamma_{4}  \\
    & - 880 \gamma_{1}^5 \gamma_{2} \gamma_{4} - 331776 \lambda^2 \gamma_0^2 \gamma_{1}^2 \gamma_{2}^2 \gamma_{4} - 82944 \lambda \gamma_0 \gamma_{1}^3 \gamma_{2}^2 \gamma_{4} - 5216 \gamma_{1}^4 \gamma_{2}^2 \gamma_{4}
    + 768 \lambda^2 \gamma_0^2 \gamma_{1}^2 \gamma_{5}^2
    \\
    & - 110592 \lambda \gamma_0 \gamma_{1}^2 \gamma_{2}^3 \gamma_{4} - 13824 \gamma_{1}^3 \gamma_{2}^3 \gamma_{4} - 13824 \gamma_{1}^2 \gamma_{2}^4 \gamma_{4} + 442368 \lambda^4 \gamma_0^4 \gamma_{3} \gamma_{4}  \\
    & + 221184 \lambda^3 \gamma_0^3 \gamma_{1} \gamma_{3} \gamma_{4} + 41472 \lambda^2 \gamma_0^2 \gamma_{1}^2 \gamma_{3} \gamma_{4} + 3456 \lambda \gamma_0 \gamma_{1}^3 \gamma_{3} \gamma_{4} + 120 \gamma_{1}^4 \gamma_{3} \gamma_{4}  \\
    & + 884736 \lambda^3 \gamma_0^3 \gamma_{2} \gamma_{3} \gamma_{4} + 331776 \lambda^2 \gamma_0^2 \gamma_{1} \gamma_{2} \gamma_{3} \gamma_{4} + 41472 \lambda \gamma_0 \gamma_{1}^2 \gamma_{2} \gamma_{3} \gamma_{4}  \\
    & + 1728 \gamma_{1}^3 \gamma_{2} \gamma_{3} \gamma_{4} + 663552 \lambda^2 \gamma_0^2 \gamma_{2}^2 \gamma_{3} \gamma_{4} + 165888 \lambda \gamma_0 \gamma_{1} \gamma_{2}^2 \gamma_{3} \gamma_{4} + 10368 \gamma_{1}^2 \gamma_{2}^2 \gamma_{3} \gamma_{4}  \\
    & + 221184 \lambda \gamma_0 \gamma_{2}^3 \gamma_{3} \gamma_{4} + 27648 \gamma_{1} \gamma_{2}^3 \gamma_{3} \gamma_{4} + 27648 \gamma_{2}^4 \gamma_{3} \gamma_{4} + 884736 \lambda^4 \gamma_0^4 \gamma_{4}^2   \\
    & + 82944 \lambda^2 \gamma_0^2 \gamma_{1}^2 \gamma_{4}^2 + 6912 \lambda \gamma_0 \gamma_{1}^3 \gamma_{4}^2 + 240 \gamma_{1}^4 \gamma_{4}^2 + 1769472 \lambda^3 \gamma_0^3 \gamma_{2} \gamma_{4}^2   \\
    & + 82944 \lambda \gamma_0 \gamma_{1}^2 \gamma_{2} \gamma_{4}^2 + 3456 \gamma_{1}^3 \gamma_{2} \gamma_{4}^2 + 1327104 \lambda^2 \gamma_0^2 \gamma_{2}^2 \gamma_{4}^2 + 331776 \lambda \gamma_0 \gamma_{1} \gamma_{2}^2 \gamma_{4}^2  \\
    & + 20736 \gamma_{1}^2 \gamma_{2}^2 \gamma_{4}^2 + 442368 \lambda \gamma_0 \gamma_{2}^3 \gamma_{4}^2 + 55296 \gamma_{1} \gamma_{2}^3 \gamma_{4}^2 + 55296 \gamma_{2}^4 \gamma_{4}^2 + 3072 \lambda^3 \gamma_0^3 \gamma_{1}^3 \gamma_{5}  \\
    & + 1024 \lambda^2 \gamma_0^2 \gamma_{1}^4 \gamma_{5} + 112 \lambda \gamma_0 \gamma_{1}^5 \gamma_{5} - 2 \gamma_{1}^6 \gamma_{5} + 3072 \lambda^2 \gamma_0^2 \gamma_{1}^3 \gamma_{2} \gamma_{5} + 640 \lambda \gamma_0 \gamma_{1}^4 \gamma_{2} \gamma_{5}  \\
    & - 16 \gamma_{1}^5 \gamma_{2} \gamma_{5} + 768 \lambda \gamma_0 \gamma_{1}^3 \gamma_{2}^2 \gamma_{5} - 32 \gamma_{1}^4 \gamma_{2}^2 \gamma_{5} + 768 \lambda^2 \gamma_0^2 \gamma_{1}^2 \gamma_{3} \gamma_{5} + 48 \gamma_{1}^4 \gamma_{4} \gamma_{5}   \\
    & + 192 \lambda \gamma_0 \gamma_{1}^3 \gamma_{3} \gamma_{5} + 24 \gamma_{1}^4 \gamma_{3} \gamma_{5} + 768 \lambda \gamma_0 \gamma_{1}^2 \gamma_{2} \gamma_{3} \gamma_{5} + 96 \gamma_{1}^3 \gamma_{2} \gamma_{3} \gamma_{5} + 192 \gamma_{1}^2 \gamma_{2}^2 \gamma_{3} \gamma_{5}  \\
    &   + 192 \lambda \gamma_0 \gamma_{1}^3 \gamma_{5}^2 + 36 \gamma_{1}^4 \gamma_{5}^2  + 768 \lambda \gamma_0 \gamma_{1}^2 \gamma_{2} \gamma_{5}^2 + 96 \gamma_{1}^3 \gamma_{2} \gamma_{5}^2 + 192 \gamma_{1}^2 \gamma_{2}^2 \gamma_{5}^2 + 1024 \lambda^2 \gamma_0^2 \gamma_{1}^4 \gamma_{6}   \\
    & + 256 \lambda \gamma_0 \gamma_{1}^5 \gamma_{6} - 8 \gamma_{1}^6 \gamma_{6} + 1024 \lambda \gamma_0 \gamma_{1}^4 \gamma_{2} \gamma_{6} - 64 \gamma_{1}^5 \gamma_{2} \gamma_{6} - 128 \gamma_{1}^4 \gamma_{2}^2 \gamma_{6} + 48 \gamma_{1}^4 \gamma_{3} \gamma_{6}  \\
    & + 192 \gamma_{1}^4 \gamma_{4} \gamma_{6} + 192 \gamma_{1}^4 \gamma_{5} \gamma_{6} + 384 \gamma_{1}^4 \gamma_{6}^2
    + 442368 \lambda^3 \gamma_0^3 \gamma_{1} \gamma_{4}^2
    + 663552 \lambda^2 \gamma_0^2 \gamma_{1} \gamma_{2} \gamma_{4}^2
    ,
    \end{split}
\end{equation}
\begin{equation}
    \begin{split}
    \xi_4 = & \,\,
    2 \lambda \gamma_0 \gamma_{1}^5 - 16 \lambda^2 \gamma_0^2 \gamma_{1}^4 + 8 \lambda \gamma_0 \gamma_{1}^4 \gamma_{2} - 32 \lambda^2 \gamma_0^2 \gamma_{1}^2 \gamma_{3} - 20 \lambda \gamma_0 \gamma_{1}^3 \gamma_{3} - 32 \lambda \gamma_0 \gamma_{1}^2 \gamma_{2} \gamma_{3}   \\
    & - 8 \gamma_{1}^2 \gamma_{2}^2 \gamma_{3} + 256 \lambda^2 \gamma_0^2 \gamma_{3}^2 + 64 \lambda \gamma_0 \gamma_{1} \gamma_{3}^2 + 2 \gamma_{1}^2 \gamma_{3}^2 + 256 \lambda \gamma_0 \gamma_{2} \gamma_{3}^2 + 32 \gamma_{1} \gamma_{2} \gamma_{3}^2 + 64 \gamma_{2}^2 \gamma_{3}^2  \\
    & - 24 \lambda \gamma_0 \gamma_{1}^3 \gamma_{5} + \gamma_{1}^4 \gamma_{5} + 4 \gamma_{1}^3 \gamma_{2} \gamma_{5} - 8 \gamma_{1}^2 \gamma_{3} \gamma_{5} - 8 \gamma_{1}^2 \gamma_{5}^2 - 2 \gamma_{1}^3 \gamma_{2} \gamma_{3}
    ,
    \end{split}
\end{equation}
\begin{equation}
    \begin{split}
    \xi_5 = & \,\,
    2560 \lambda^4 \gamma_0^4 \gamma_{1}^4 + 128 \lambda^3 \gamma_0^3 \gamma_{1}^5 - 40 \lambda^2 \gamma_0^2 \gamma_{1}^6 + 4 \lambda \gamma_0 \gamma_{1}^7 + 512 \lambda^3 \gamma_0^3 \gamma_{1}^4 \gamma_{2} - 320 \lambda^2 \gamma_0^2 \gamma_{1}^5 \gamma_{2}  \\
    & + 48 \lambda \gamma_0 \gamma_{1}^6 \gamma_{2} - 640 \lambda^2 \gamma_0^2 \gamma_{1}^4 \gamma_{2}^2 + 192 \lambda \gamma_0 \gamma_{1}^5 \gamma_{2}^2 + 256 \lambda \gamma_0 \gamma_{1}^4 \gamma_{2}^3 + 70656 \lambda^4 \gamma_0^4 \gamma_{1}^2 \gamma_{3}  \\
    & + 37632 \lambda^3 \gamma_0^3 \gamma_{1}^3 \gamma_{3} + 6992 \lambda^2 \gamma_0^2 \gamma_{1}^4 \gamma_{3} + 536 \lambda \gamma_0 \gamma_{1}^5 \gamma_{3} + 17 \gamma_{1}^6 \gamma_{3} + 141312 \lambda^3 \gamma_0^3 \gamma_{1}^2 \gamma_{2} \gamma_{3}  \\
    & + 54912 \lambda^2 \gamma_0^2 \gamma_{1}^3 \gamma_{2} \gamma_{3} + 6608 \lambda \gamma_0 \gamma_{1}^4 \gamma_{2} \gamma_{3} + 268 \gamma_{1}^5 \gamma_{2} \gamma_{3} + 105984 \lambda^2 \gamma_0^2 \gamma_{1}^2 \gamma_{2}^2 \gamma_{3}  \\
    & + 26688 \lambda \gamma_0 \gamma_{1}^3 \gamma_{2}^2 \gamma_{3} + 1604 \gamma_{1}^4 \gamma_{2}^2 \gamma_{3} + 35328 \lambda \gamma_0 \gamma_{1}^2 \gamma_{2}^3 \gamma_{3} + 4320 \gamma_{1}^3 \gamma_{2}^3 \gamma_{3}  \\
    & + 4416 \gamma_{1}^2 \gamma_{2}^4 \gamma_{3} - 159744 \lambda^4 \gamma_0^4 \gamma_{3}^2 - 79872 \lambda^3 \gamma_0^3 \gamma_{1} \gamma_{3}^2 - 14592 \lambda^2 \gamma_0^2 \gamma_{1}^2 \gamma_{3}^2 - 1152 \lambda \gamma_0 \gamma_{1}^3 \gamma_{3}^2  \\
    & - 36 \gamma_{1}^4 \gamma_{3}^2 - 319488 \lambda^3 \gamma_0^3 \gamma_{2} \gamma_{3}^2 - 119808 \lambda^2 \gamma_0^2 \gamma_{1} \gamma_{2} \gamma_{3}^2 - 14592 \lambda \gamma_0 \gamma_{1}^2 \gamma_{2} \gamma_{3}^2  \\
    & - 576 \gamma_{1}^3 \gamma_{2} \gamma_{3}^2 - 239616 \lambda^2 \gamma_0^2 \gamma_{2}^2 \gamma_{3}^2 - 59904 \lambda \gamma_0 \gamma_{1} \gamma_{2}^2 \gamma_{3}^2 - 3648 \gamma_{1}^2 \gamma_{2}^2 \gamma_{3}^2  \\
    & - 79872 \lambda \gamma_0 \gamma_{2}^3 \gamma_{3}^2 - 9984 \gamma_{1} \gamma_{2}^3 \gamma_{3}^2 - 9984 \gamma_{2}^4 \gamma_{3}^2 + 258048 \lambda^4 \gamma_0^4 \gamma_{1}^2 \gamma_{4} + 129024 \lambda^3 \gamma_0^3 \gamma_{1}^3 \gamma_{4}  \\
    & + 23744 \lambda^2 \gamma_0^2 \gamma_{1}^4 \gamma_{4} + 1904 \lambda \gamma_0 \gamma_{1}^5 \gamma_{4} + 68 \gamma_{1}^6 \gamma_{4} + 516096 \lambda^3 \gamma_0^3 \gamma_{1}^2 \gamma_{2} \gamma_{4}  \\
    & + 193536 \lambda^2 \gamma_0^2 \gamma_{1}^3 \gamma_{2} \gamma_{4} + 23744 \lambda \gamma_0 \gamma_{1}^4 \gamma_{2} \gamma_{4} + 1048 \gamma_{1}^5 \gamma_{2} \gamma_{4} + 387072 \lambda^2 \gamma_0^2 \gamma_{1}^2 \gamma_{2}^2 \gamma_{4}  \\
    & + 96768 \lambda \gamma_0 \gamma_{1}^3 \gamma_{2}^2 \gamma_{4} + 6128 \gamma_{1}^4 \gamma_{2}^2 \gamma_{4} + 129024 \lambda \gamma_0 \gamma_{1}^2 \gamma_{2}^3 \gamma_{4} + 16128 \gamma_{1}^3 \gamma_{2}^3 \gamma_{4}  \\
    & + 16128 \gamma_{1}^2 \gamma_{2}^4 \gamma_{4} - 884736 \lambda^4 \gamma_0^4 \gamma_{3} \gamma_{4} - 442368 \lambda^3 \gamma_0^3 \gamma_{1} \gamma_{3} \gamma_{4} - 82944 \lambda^2 \gamma_0^2 \gamma_{1}^2 \gamma_{3} \gamma_{4}  \\
    & - 6912 \lambda \gamma_0 \gamma_{1}^3 \gamma_{3} \gamma_{4} - 240 \gamma_{1}^4 \gamma_{3} \gamma_{4} - 1769472 \lambda^3 \gamma_0^3 \gamma_{2} \gamma_{3} \gamma_{4} - 663552 \lambda^2 \gamma_0^2 \gamma_{1} \gamma_{2} \gamma_{3} \gamma_{4}  \\
    & - 82944 \lambda \gamma_0 \gamma_{1}^2 \gamma_{2} \gamma_{3} \gamma_{4} - 3456 \gamma_{1}^3 \gamma_{2} \gamma_{3} \gamma_{4} - 1327104 \lambda^2 \gamma_0^2 \gamma_{2}^2 \gamma_{3} \gamma_{4} + 384 \lambda \gamma_0 \gamma_{1}^3 \gamma_{3} \gamma_{5} \\
    & - 331776 \lambda \gamma_0 \gamma_{1} \gamma_{2}^2 \gamma_{3} \gamma_{4} - 20736 \gamma_{1}^2 \gamma_{2}^2 \gamma_{3} \gamma_{4} - 442368 \lambda \gamma_0 \gamma_{2}^3 \gamma_{3} \gamma_{4} + 1536 \lambda^2 \gamma_0^2 \gamma_{1}^2 \gamma_{3} \gamma_{5} \\
    & - 55296 \gamma_{1} \gamma_{2}^3 \gamma_{3} \gamma_{4} - 55296 \gamma_{2}^4 \gamma_{3} \gamma_{4} - 1769472 \lambda^4 \gamma_0^4 \gamma_{4}^2 - 884736 \lambda^3 \gamma_0^3 \gamma_{1} \gamma_{4}^2  \\
    & - 165888 \lambda^2 \gamma_0^2 \gamma_{1}^2 \gamma_{4}^2 - 13824 \lambda \gamma_0 \gamma_{1}^3 \gamma_{4}^2 - 480 \gamma_{1}^4 \gamma_{4}^2 - 3538944 \lambda^3 \gamma_0^3 \gamma_{2} \gamma_{4}^2  \\
    & - 1327104 \lambda^2 \gamma_0^2 \gamma_{1} \gamma_{2} \gamma_{4}^2 - 165888 \lambda \gamma_0 \gamma_{1}^2 \gamma_{2} \gamma_{4}^2 - 6912 \gamma_{1}^3 \gamma_{2} \gamma_{4}^2 - 2654208 \lambda^2 \gamma_0^2 \gamma_{2}^2 \gamma_{4}^2  \\
    & - 663552 \lambda \gamma_0 \gamma_{1} \gamma_{2}^2 \gamma_{4}^2 - 41472 \gamma_{1}^2 \gamma_{2}^2 \gamma_{4}^2 - 884736 \lambda \gamma_0 \gamma_{2}^3 \gamma_{4}^2 - 110592 \gamma_{1} \gamma_{2}^3 \gamma_{4}^2  \\
    & - 110592 \gamma_{2}^4 \gamma_{4}^2 + 4608 \lambda^3 \gamma_0^3 \gamma_{1}^3 \gamma_{5} + 512 \lambda^2 \gamma_0^2 \gamma_{1}^4 \gamma_{5} - 88 \lambda \gamma_0 \gamma_{1}^5 \gamma_{5} + 2 \gamma_{1}^6 \gamma_{5}  \\
    & + 3840 \lambda^2 \gamma_0^2 \gamma_{1}^3 \gamma_{2} \gamma_{5} - 256 \lambda \gamma_0 \gamma_{1}^4 \gamma_{2} \gamma_{5} + 4 \gamma_{1}^5 \gamma_{2} \gamma_{5} + 384 \lambda \gamma_0 \gamma_{1}^3 \gamma_{2}^2 \gamma_{5} - 64 \gamma_{1}^4 \gamma_{2}^2 \gamma_{5}  \\
    & + 1536 \lambda \gamma_0 \gamma_{1}^2 \gamma_{2} \gamma_{3} \gamma_{5} + 192 \gamma_{1}^3 \gamma_{2} \gamma_{3} \gamma_{5} + 384 \gamma_{1}^2 \gamma_{2}^2 \gamma_{3} \gamma_{5} - 96 \gamma_{1}^4 \gamma_{4} \gamma_{5} - 192 \gamma_{1}^3 \gamma_{2}^3 \gamma_{5}  \\
    & + 1536 \lambda^2 \gamma_0^2 \gamma_{1}^2 \gamma_{5}^2 + 384 \lambda \gamma_0 \gamma_{1}^3 \gamma_{5}^2 - 24 \gamma_{1}^4 \gamma_{5}^2 + 1536 \lambda \gamma_0 \gamma_{1}^2 \gamma_{2} \gamma_{5}^2 + 192 \gamma_{1}^3 \gamma_{2} \gamma_{5}^2  \\
    & + 384 \gamma_{1}^2 \gamma_{2}^2 \gamma_{5}^2 - 1792 \lambda^2 \gamma_0^2 \gamma_{1}^4 \gamma_{6} - 448 \lambda \gamma_0 \gamma_{1}^5 \gamma_{6} + 20 \gamma_{1}^6 \gamma_{6} - 1792 \lambda \gamma_0 \gamma_{1}^4 \gamma_{2} \gamma_{6}  \\
    & + 160 \gamma_{1}^5 \gamma_{2} \gamma_{6} + 320 \gamma_{1}^4 \gamma_{2}^2 \gamma_{6} - 96 \gamma_{1}^4 \gamma_{3} \gamma_{6} - 384 \gamma_{1}^4 \gamma_{4} \gamma_{6} - 384 \gamma_{1}^4 \gamma_{5} \gamma_{6}
     - 768 \gamma_{1}^4 \gamma_{6}^2
     ,
    \end{split}
\end{equation}
\begin{equation}
    \begin{split} \xi_6 = & \,\,
    -384 \lambda^3 \gamma_0^3 \gamma_{1}^5 - 88 \lambda^2 \gamma_0^2 \gamma_{1}^6 - 4 \lambda \gamma_0 \gamma_{1}^7 - 1536 \lambda^3 \gamma_0^3 \gamma_{1}^4 \gamma_{2} - 704 \lambda^2 \gamma_0^2 \gamma_{1}^5 \gamma_{2} - 48 \lambda \gamma_0 \gamma_{1}^6 \gamma_{2}  \\
    & - 1408 \lambda^2 \gamma_0^2 \gamma_{1}^4 \gamma_{2}^2 - 192 \lambda \gamma_0 \gamma_{1}^5 \gamma_{2}^2 - 256 \lambda \gamma_0 \gamma_{1}^4 \gamma_{2}^3 + 9216 \lambda^4 \gamma_0^4 \gamma_{1}^2 \gamma_{3} + 5376 \lambda^3 \gamma_0^3 \gamma_{1}^3 \gamma_{3}  \\
    & + 944 \lambda^2 \gamma_0^2 \gamma_{1}^4 \gamma_{3} + 56 \lambda \gamma_0 \gamma_{1}^5 \gamma_{3} + \gamma_{1}^6 \gamma_{3} + 18432 \lambda^3 \gamma_0^3 \gamma_{1}^2 \gamma_{2} \gamma_{3} + 7296 \lambda^2 \gamma_0^2 \gamma_{1}^3 \gamma_{2} \gamma_{3}  \\
    & + 752 \lambda \gamma_0 \gamma_{1}^4 \gamma_{2} \gamma_{3} + 20 \gamma_{1}^5 \gamma_{2} \gamma_{3} + 13824 \lambda^2 \gamma_0^2 \gamma_{1}^2 \gamma_{2}^2 \gamma_{3} + 3264 \lambda \gamma_0 \gamma_{1}^3 \gamma_{2}^2 \gamma_{3}  \\
    & + 148 \gamma_{1}^4 \gamma_{2}^2 \gamma_{3} + 4608 \lambda \gamma_0 \gamma_{1}^2 \gamma_{2}^3 \gamma_{3} + 480 \gamma_{1}^3 \gamma_{2}^3 \gamma_{3} + 576 \gamma_{1}^2 \gamma_{2}^4 \gamma_{3} - 43008 \lambda^4 \gamma_0^4 \gamma_{3}^2  \\
    & - 21504 \lambda^3 \gamma_0^3 \gamma_{1} \gamma_{3}^2 - 3840 \lambda^2 \gamma_0^2 \gamma_{1}^2 \gamma_{3}^2 - 288 \lambda \gamma_0 \gamma_{1}^3 \gamma_{3}^2 - 8 \gamma_{1}^4 \gamma_{3}^2 - 86016 \lambda^3 \gamma_0^3 \gamma_{2} \gamma_{3}^2  \\
    & - 32256 \lambda^2 \gamma_0^2 \gamma_{1} \gamma_{2} \gamma_{3}^2 - 3840 \lambda \gamma_0 \gamma_{1}^2 \gamma_{2} \gamma_{3}^2 - 144 \gamma_{1}^3 \gamma_{2} \gamma_{3}^2 - 64512 \lambda^2 \gamma_0^2 \gamma_{2}^2 \gamma_{3}^2  \\
    & - 16128 \lambda \gamma_0 \gamma_{1} \gamma_{2}^2 \gamma_{3}^2 - 960 \gamma_{1}^2 \gamma_{2}^2 \gamma_{3}^2 - 21504 \lambda \gamma_0 \gamma_{2}^3 \gamma_{3}^2 - 2688 \gamma_{1} \gamma_{2}^3 \gamma_{3}^2 - 2688 \gamma_{2}^4 \gamma_{3}^2  \\
    & + 12288 \lambda^4 \gamma_0^4 \gamma_{1}^2 \gamma_{4} + 6144 \lambda^3 \gamma_0^3 \gamma_{1}^3 \gamma_{4} + 1088 \lambda^2 \gamma_0^2 \gamma_{1}^4 \gamma_{4} + 80 \lambda \gamma_0 \gamma_{1}^5 \gamma_{4} + 4 \gamma_{1}^6 \gamma_{4}  \\
    & + 24576 \lambda^3 \gamma_0^3 \gamma_{1}^2 \gamma_{2} \gamma_{4} + 9216 \lambda^2 \gamma_0^2 \gamma_{1}^3 \gamma_{2} \gamma_{4} + 1088 \lambda \gamma_0 \gamma_{1}^4 \gamma_{2} \gamma_{4} + 56 \gamma_{1}^5 \gamma_{2} \gamma_{4}  \\
    & + 18432 \lambda^2 \gamma_0^2 \gamma_{1}^2 \gamma_{2}^2 \gamma_{4} + 4608 \lambda \gamma_0 \gamma_{1}^3 \gamma_{2}^2 \gamma_{4} + 304 \gamma_{1}^4 \gamma_{2}^2 \gamma_{4} + 6144 \lambda \gamma_0 \gamma_{1}^2 \gamma_{2}^3 \gamma_{4}  \\
    & + 768 \gamma_{1}^3 \gamma_{2}^3 \gamma_{4} + 768 \gamma_{1}^2 \gamma_{2}^4 \gamma_{4} - 147456 \lambda^4 \gamma_0^4 \gamma_{3} \gamma_{4} - 73728 \lambda^3 \gamma_0^3 \gamma_{1} \gamma_{3} \gamma_{4}  \\
    & - 13824 \lambda^2 \gamma_0^2 \gamma_{1}^2 \gamma_{3} \gamma_{4} - 1152 \lambda \gamma_0 \gamma_{1}^3 \gamma_{3} \gamma_{4} - 40 \gamma_{1}^4 \gamma_{3} \gamma_{4} - 294912 \lambda^3 \gamma_0^3 \gamma_{2} \gamma_{3} \gamma_{4}  \\
    & - 110592 \lambda^2 \gamma_0^2 \gamma_{1} \gamma_{2} \gamma_{3} \gamma_{4} - 13824 \lambda \gamma_0 \gamma_{1}^2 \gamma_{2} \gamma_{3} \gamma_{4} - 576 \gamma_{1}^3 \gamma_{2} \gamma_{3} \gamma_{4}  \\
    & - 221184 \lambda^2 \gamma_0^2 \gamma_{2}^2 \gamma_{3} \gamma_{4} - 55296 \lambda \gamma_0 \gamma_{1} \gamma_{2}^2 \gamma_{3} \gamma_{4} - 3456 \gamma_{1}^2 \gamma_{2}^2 \gamma_{3} \gamma_{4} - 73728 \lambda \gamma_0 \gamma_{2}^3 \gamma_{3} \gamma_{4}  \\
    & - 9216 \gamma_{1} \gamma_{2}^3 \gamma_{3} \gamma_{4} - 9216 \gamma_{2}^4 \gamma_{3} \gamma_{4} - 294912 \lambda^4 \gamma_0^4 \gamma_{4}^2 - 147456 \lambda^3 \gamma_0^3 \gamma_{1} \gamma_{4}^2 - 27648 \lambda^2 \gamma_0^2 \gamma_{1}^2 \gamma_{4}^2  \\
    & - 2304 \lambda \gamma_0 \gamma_{1}^3 \gamma_{4}^2 - 80 \gamma_{1}^4 \gamma_{4}^2 - 589824 \lambda^3 \gamma_0^3 \gamma_{2} \gamma_{4}^2 - 221184 \lambda^2 \gamma_0^2 \gamma_{1} \gamma_{2} \gamma_{4}^2 - 27648 \lambda \gamma_0 \gamma_{1}^2 \gamma_{2} \gamma_{4}^2  \\
    & - 1152 \gamma_{1}^3 \gamma_{2} \gamma_{4}^2 - 442368 \lambda^2 \gamma_0^2 \gamma_{2}^2 \gamma_{4}^2 - 110592 \lambda \gamma_0 \gamma_{1} \gamma_{2}^2 \gamma_{4}^2 - 6912 \gamma_{1}^2 \gamma_{2}^2 \gamma_{4}^2  \\
    & - 147456 \lambda \gamma_0 \gamma_{2}^3 \gamma_{4}^2 - 18432 \gamma_{1} \gamma_{2}^3 \gamma_{4}^2 - 18432 \gamma_{2}^4 \gamma_{4}^2 + 1536 \lambda^3 \gamma_0^3 \gamma_{1}^3 \gamma_{5} + 128 \lambda^2 \gamma_0^2 \gamma_{1}^4 \gamma_{5}  \\
    & - 40 \lambda \gamma_0 \gamma_{1}^5 \gamma_{5} - 2 \gamma_{1}^6 \gamma_{5} + 768 \lambda^2 \gamma_0^2 \gamma_{1}^3 \gamma_{2} \gamma_{5} - 256 \lambda \gamma_0 \gamma_{1}^4 \gamma_{2} \gamma_{5} - 28 \gamma_{1}^5 \gamma_{2} \gamma_{5}  \\
    & - 384 \lambda \gamma_0 \gamma_{1}^3 \gamma_{2}^2 \gamma_{5} - 128 \gamma_{1}^4 \gamma_{2}^2 \gamma_{5} - 192 \gamma_{1}^3 \gamma_{2}^3 \gamma_{5} + 768 \lambda^2 \gamma_0^2 \gamma_{1}^2 \gamma_{3} \gamma_{5}  \\
    & + 192 \lambda \gamma_0 \gamma_{1}^3 \gamma_{3} \gamma_{5} + 8 \gamma_{1}^4 \gamma_{3} \gamma_{5} + 768 \lambda \gamma_0 \gamma_{1}^2 \gamma_{2} \gamma_{3} \gamma_{5} + 96 \gamma_{1}^3 \gamma_{2} \gamma_{3} \gamma_{5}  \\
    & + 192 \gamma_{1}^2 \gamma_{2}^2 \gamma_{3} \gamma_{5} - 16 \gamma_{1}^4 \gamma_{4} \gamma_{5} + 768 \lambda^2 \gamma_0^2 \gamma_{1}^2 \gamma_{5}^2 + 192 \lambda \gamma_0 \gamma_{1}^3 \gamma_{5}^2 + 4 \gamma_{1}^4 \gamma_{5}^2  \\
    & + 768 \lambda \gamma_0 \gamma_{1}^2 \gamma_{2} \gamma_{5}^2 + 96 \gamma_{1}^3 \gamma_{2} \gamma_{5}^2 + 192 \gamma_{1}^2 \gamma_{2}^2 \gamma_{5}^2 - 256 \lambda^2 \gamma_0^2 \gamma_{1}^4 \gamma_{6} - 64 \lambda \gamma_0 \gamma_{1}^5 \gamma_{6}  \\
    & + 4 \gamma_{1}^6 \gamma_{6} - 256 \lambda \gamma_0 \gamma_{1}^4 \gamma_{2} \gamma_{6} + 32 \gamma_{1}^5 \gamma_{2} \gamma_{6} + 64 \gamma_{1}^4 \gamma_{2}^2 \gamma_{6} - 16 \gamma_{1}^4 \gamma_{3} \gamma_{6}  \\
    & - 64 \gamma_{1}^4 \gamma_{4} \gamma_{6} - 64 \gamma_{1}^4 \gamma_{5} \gamma_{6} - 128 \gamma_{1}^4 \gamma_{6}^2    .
    \end{split}
\end{equation}
The parameters $\ga_{0,1,2,...,6}$ in these expressions
represent the ambiguity in the parametrization of quantum metric~\eq{gen-bfm}.
Let us note that these bulky formulas, anyway, admit a completely
invariant result when used in the potential (\ref{potential_final}),
providing the very compact Eq.~(\ref{NewNew}).

\newpage
\section*{Appendix B}
\label{ApB}

Here we derive the equation DM(35). All
equations like DM(35) are numbered according to
Ref.~\cite{DM97}. For comparison reasons,
the notations in this section
also follows~\cite{DM97}. The principal difference with the main text
conventions is the opposite sign for the metric signature.

The starting expression is
\beq
\De S_{mM} = -\frac{1}{32\pi^2} \int d^4y \sqrt{-g}
\,m_{\mu\nu\rho\si}\,\ln (-\cx) \,M^{\rho\si\mu\nu},
\label{35}
\eeq
where DM(15) gives
\beq
&&
M^{\rho\si\mu\nu} = \frac{M\ka^2}{8}\int d\tau'
\de^4(y-x(\tau'))\,
\big(2{\dot x}^\rho{\dot x}^\si{\dot x}^\mu{\dot x}^\nu
+ {\dot x}^\rho{\dot x}^\si g^{\mu\nu} \big),
\nn
\\
&&
m_{\mu\nu\rho\si}  = \frac{m\ka^2}{8}\int d\tau
\de^4(y-z(\tau))\,\big(
2{\dot z}_\rho{\dot z}_\si{\dot z}_\mu{\dot z}_\nu
+ {\dot z}_\mu{\dot z}_\nu g_{\rho\si} \big) .
\label{mM}
\eeq
Hereafter, ${\dot x}_\mu = \frac{d x_\mu}{d\tau'}$ and
${\dot z}_\mu = \frac{d z_\mu}{d\tau}$.
Before substituting (\ref{mM}) into (\ref{35}), we take into
account that the massive source $M$ is static and situated at
the origin, such that $x^0=\tau$ and $x^k \equiv 0$. Furthermore,
\beq
\label{mv}
g^{\mu\nu} {\dot z}_\mu{\dot z}_\nu  = - 1.
\eeq
Then, the contractions in (\ref{35}) give
\beq
\De S_{mM}
&=& - \frac{mM\ka^4}{64 \cdot 32\pi^2}
 \int dy^0\,d^3{\vec y} \sqrt{-g (y)}
 \int d\tau\, \de^4(y - z(\tau))
 \big( 2{\dot z}_\rho{\dot z}_\si{\dot z}_\mu{\dot z}_\nu
 + {\dot z}_\rho{\dot z}_\si g_{\mu\nu} \big)
\nn
\\
&&
\quad
\times \,
 \,\ln (-\cx) \,
\int d \tau' \,\de (y^0 - \tau') \de^3 ({\vec y})
\big(  2{\de}^\rho_0 {\de}^\si_0 {\de}^\mu_0 {\de}^\nu_0
+ {\de}^\rho_0{\de}^\si_0 g^{\rho\si} \big)
\nn
\\
&=&
\quad
- \frac{mM\ka^4}{64 \cdot 32\pi^2}
\int dy^0\,d^3{\vec y} \sqrt{-g (y)}
\int d\tau\,\,
\de (y^0 - z^0(\tau)) \de^3 ({\vec y}-{\vec z}(\tau))
\nn
\\
&&
\quad
\times
\big(
     g_{00}g^{\rho\si}{\dot z}_\rho{\dot z}_\si
+  2 {\dot z}_0{\dot z}_0 g^{\rho\si}{\dot z}_\rho{\dot z}_\si
+  2 g_{00} {\dot z}_0{\dot z}_0
+ 4{\dot z}_0 {\dot z}_0 {\dot z}_0 {\dot z}_0
\big)
\nn
\\
&&
\quad
\times
 \,\ln (-\cx) \,
 \int d\tau'\, \de (y^0 - \tau')\de^3({\vec y}).
\label{35a}
\eeq
Taking into account \eq{mv},
the formula (\ref{35a}) boils down to
\beq
\De S_{mM}
&=&
- \frac{mM\ka^4}{64 \cdot 32\pi^2}
\int dy^0\,d^3{\vec y} \sqrt{-g (y)}
\int d\tau\, \de (y^0 - z^0) \de^3 ({\vec y}-{\vec z}(\tau))
\nn
\\
&&
\quad
\times
\big(
-  g_{00}
-   2 {\dot z}_0{\dot z}_0
+  2 g_{00} {\dot z}_0{\dot z}_0
+ 4 {\dot z}_0{\dot z}_0 {\dot z}_0{\dot z}_0
\big)
\,\ln (-\cx) \,
\de^3({\vec y}).
\label{35b}
\eeq
Taking the integral over $y^0$ and changing the order of integration,
we get $y^0 \to \tau$. Similarly, taking $d^3{\vec y}$ results in
${\vec y}\to {\vec z}(\tau)$.
Assuming non relativistic $z^\mu$, i.e., $z^0=\tau$ and ${\dot z}_k=0$,
we get
\beq
\De S_{mM}
&=& - \frac{mM\ka^4}{64 \cdot 32\pi^2}
\int d\tau
\big(
-  g_{00}
-   2 {\dot z}_0{\dot z}_0
-  2 {\dot z}_0{\dot z}_0
+ 4 {\dot z}_0{\dot z}_0 {\dot z}_0{\dot z}_0
\big)
 \,\ln (-\cx) \, \de^3 ({\vec z (\tau)})
 \nn
 \\
 &&
 \quad
 = - \frac{mM\ka^4}{512\pi^2}
\int d\tau\,\,\frac14\,
 \,\ln (-\cx) \, \de^3 ({\vec z (\tau)}) ,
\label{35d}
\eeq
that is four time more than DM(35).
Correcting the missing
factor of 4, the coefficient in DM(43) is $-1/6$ instead of $7/12$.
This gives a Newtonian potential in the form \eq{NewNew}.


\end{document}